\documentclass[aps,twocolumn,superscriptaddress,showpacs]{revtex4-1}
\usepackage[pdftex]{graphicx}
\usepackage{graphicx}
\usepackage{dcolumn}
\usepackage{bm}
\usepackage{amssymb}
\makeatletter
\def\captionof#1#2{{\def\@captype{#1}#2}}
\makeatother

\usepackage[english]{babel}
\usepackage{times}
\usepackage{graphicx}
\usepackage{amsfonts}
\usepackage{psfrag}
\usepackage{verbatim}
\usepackage{color}

\begin{document}

\title{Defects dynamics following thermal quenches in square spin-ice}
\author{Demian Levis}
\affiliation{Universit\'e  Pierre et Marie Curie - Paris 6, Laboratoire de Physique Th\'eorique et Hautes Energies,
4, Place Jussieu, Tour 13, 5\`eme \'etage, 75252 Paris Cedex 05, France}
\affiliation{Laboratoire Charles Coulomb, UMR 5221 CNRS and Universit\'e Montpellier 2, Montpellier, France}
\author{Leticia F. Cugliandolo}
\affiliation{Universit\'e Pierre et Marie Curie - Paris 6, Laboratoire de Physique Th\'eorique et Hautes Energies,
4, Place Jussieu, Tour 13, 5\`eme \'etage, 75252 Paris Cedex 05, France}


\begin{abstract}
We present a study of the single spin flip stochastic dynamics of the 
two dimensional sixteen vertex model. We single out  several dynamic regimes
controlled by different processes that we describe. We analyse the emergence of very long-lived 
metastable states and their dependence on the system size, boundary conditions, and 
working parameters. We investigate the coarsening process after quenches into the ordered
ferromagnetic and antiferromagnetic phases. We discuss our results in the context
of artificial spin ice on square lattices.
\end{abstract}
\pacs{75.10.Hk, 05.70.Ln,  75.40.Lk, 75.40.Mg }
\keywords{Suggested keywords}

\maketitle

\setlength{\textfloatsep}{10pt} 
\setlength{\intextsep}{10pt}

\tableofcontents

\section{Introduction}

Frustrated magnets are classical and quantum systems in which the interactions in combination with the lattice structure impede the spins to order in an 
optimal configuration at zero temperature~\cite{Balents2010}. In classical instances, the local minimisation of the interaction energy on a 
frustrated unit gives rise to a macroscopic degeneracy of the ground state. This occurs in spin-ice samples in which the spin interactions mimic the 
frustration of proton positions in water ice. The theoretical interest in these systems has been boosted in recent years by the artificial design of materials. 
For instance, regular arrays of elongated single-domain ferromagnetic nano-islands arranged along the sides of a $2D$ square lattice, were manufactured. 
The beauty of artificial spin-ice (ASI) is that the state of a single degree of freedom can be directly visualised with magnetic force microscopy. Moreover, 
advances in lithography allow great flexibility in their design, and the interaction parameters can be precisely controlled by tuning the island length, the 
distance between them, and the height between layers, to select the phase into which the system should set in~\cite{Wang2006,Nisoli2007,Morgan2011}. 
One of the main goals of the research on artificial spin-ices is to develop new materials that could improve the performance of data storage and data processing 
devices~\cite{Cowburn-Welland,Bader}. 

One can model spin-ice materials by taking into account dipolar interactions~\cite{Harris1997,DiepBookCH7,Moller2006,Wysin2013}
or by using a simpler vertex model~\cite{Nisoli2007,Levis2012,Foini2012}. In perfect spin-ice samples the total spin 
surrounding a lattice point is constrained to vanish according to the two-in/two-out rule, and the vertex model is integrable. For the model 
defined on a square lattice two ordered and a critical 
disordered  (spin liquid) phase have been found with powerful analytic tools~\cite{BaxterBook,LiebWuBook}.
The model with four-in or four-out arrows can also be solved
analytically and the critical character of the disordered phase is lost in this case~\cite{Baxter1971}.
Vertices that do not satisfy the two-in/two-out rule and  carry dipolar moment break 
integrability and no exact tool exists to solve models with them. We recently solved the statics of the sixteen-vertex 
(all possible states of four arrows attached to a central site) model with an extension of the Bethe-Peierls or cavity 
method~\cite{Foini2012} and we found intriguing relations between this technique and the ones of 
integrable systems. 

In this work, we will be interested in characterising the {\it single spin flip stochastic dynamics} of the sixteen vertex model on a two dimensional 
square lattice with energetically unfavored defects. Indeed, 
classical natural frustrated magnets are subject to thermal fluctuations, and these can be captured by 
vertex models coupled to a heat-bath~\cite{Levis2012,Budrikis2012}. In a previous Letter we showed that  their stochastic dynamics display 
metastability in the disordered phase, coarsening of stripes in the ferromagnetic phase, and growth of domains, that can be made isotropic, 
in the antiferromagnetic phase~\cite{Levis2012}.  Here, we extend  this analysis in several directions to be described in detail in the main text.
We work with system sizes $L = O(10^2)$  that are of the same order as the ones manufactured with lithography techniques when 
preparing artificial spin-ice samples. Finite size effects observed in the simulations are then relevant to observations in these materials.

The manuscript is organised as follows: we first present the model and the simulation method in Section~\ref{sec:ModelMethods}; next, in 
Section~\ref{sec:Metastability}, we discuss the possibility 
to set the system in metastable states with  constant number of vertices 
of each kind and an excess of defects following different quenches into the PM (Section~\ref{sec:QuenchPM}), 
$a$-FM (Section~\ref{sec:QuenchFM}) and $c$-AF 
(Section~\ref{sec:QuenchAF}) phases. In Section~\ref{sec:Coarsening} we report our results on the coarsening dynamics occurring in the 
model following the before mentioned procedure into the $a$-FM phase (Section~\ref{subsec:coarsening-FM}) and $c$-AF phase 
(Section~\ref{subsec:coarsening-AF}). Finally,
in Sect.~\ref{sec:conclusions} we present our conclusions.

\section{Model and methods}\label{sec:ModelMethods}

In this Section we present the model and the numerical techniques used in this work.

\subsection{The sixteen-vertex model}

\begin{figure}[h]
\centering
\includegraphics[scale=.25]{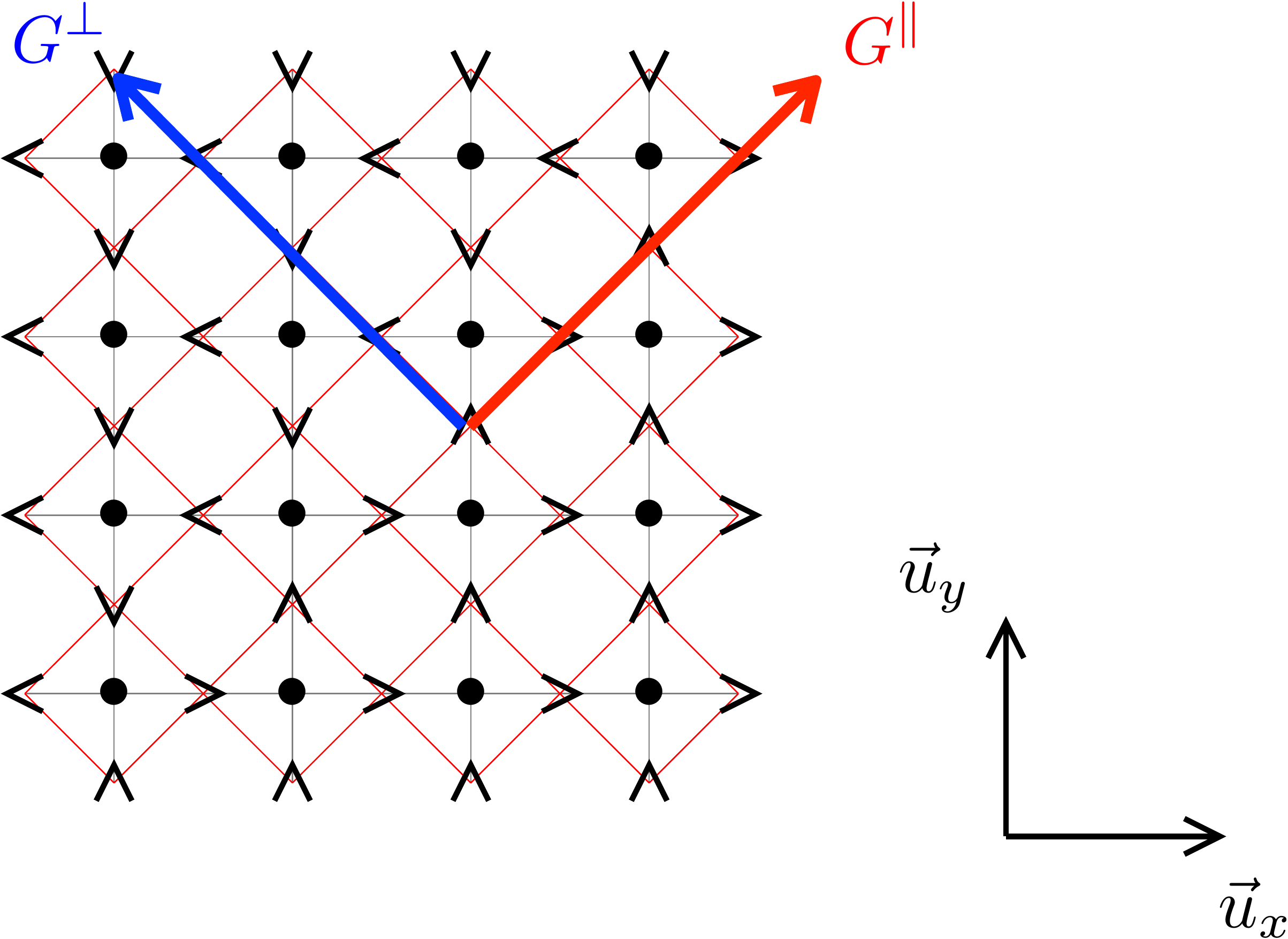} 
\caption{(Colour online.) The original lattice $\mathcal{V}$, with $L^2$ vertices, is shown in grey. Its medial lattice, $\hat{\mathcal{V}}$, with
$2L^2$ spins is shown in red. 
The Cartesian system of coordinates of the original lattice $(\vec u_x, \vec u_y)$ is shown in the right bottom end and the 
($\pi/4$)-rotated Cartesian system of coordinates used to compute parallel and perpendicular correlation functions 
is displayed on the lattice.}
\label{fig:CorrLattice}
\end{figure}

\begin{figure}[h]
\centering
\includegraphics[scale=0.76]{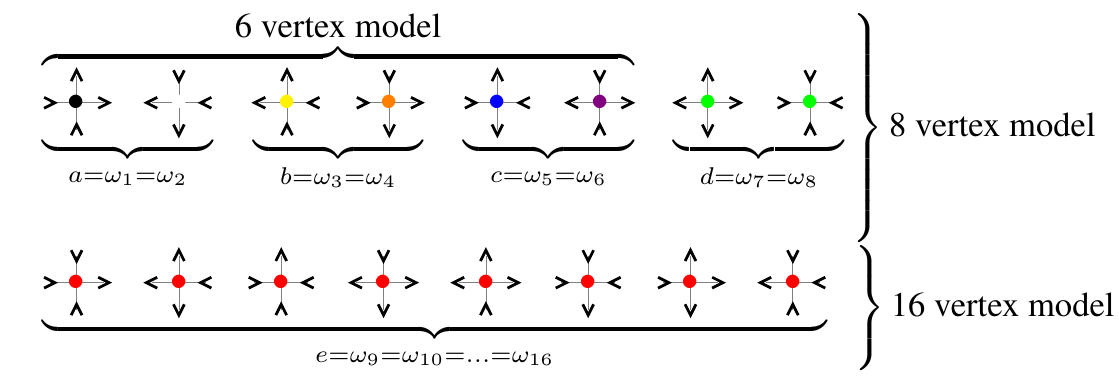} 
\caption{(Colour online.) The sixteen vertex configurations on the $2D$ square lattice 
and their  weights. The first six vertices verify the ice-rule.  The next pair
(2-fold defects)  completes the eight-vertex model and have charge $q=\pm 2$. The last eight vertices (1-fold 
defects) have charge $q=\pm 1$.}
\label{vertex_configurations}
\end{figure}

Conventional vertex models are defined on a finite dimensional lattice, typically a square one.
The degrees of freedom (Ising spins,
$q-$valued Potts variables, etc.) sit on the edges of the lattice. 
In our case
we use an $L\times L$ bidimensional  square lattice $\mathcal{V}$ with unit spacing and periodic boundary conditions (PBC). 
The midpoints of the edges of the lattice ${\cal V}$ constitute another square lattice, the medial lattice $\hat {\cal V}$.  
In the following we label the sites of $\hat{\cal V}$ with $(i,j)$, see Fig.~\ref{fig:CorrLattice}. 
In the model we focus on, the degrees 
of freedom are arrows aligned along the edges of the original square lattice ${\cal V}$. We can think of them as
Ising spins $s_{ij}=\pm 1$. 
Without loss of generality, we choose a convention such that $s = + 1$          
corresponds to an arrow pointing in the right or up direction, depending on the orientation of the link; conversely, 
$s= - 1$ corresponds to arrows pointing down or left.
When the spins joining at each vertex of ${\cal V}$ are constrained to satisfy the  two-in/two-out 
rule~\cite{Bernal1933,Pauling1935}, as for the first six vertices
in Fig.~\ref{vertex_configurations}, one has the {\it six-vertex} model. When the next two vertices in this
figure are also allowed (four-in and four-out arrows) one has the {\it eight-vertex} model. Otherwise,
the model is generalized to allow for all possible vertices with four legs and  
becomes the {\it sixteen-vertex} one, that we have already studied in~\cite{Levis2012, Foini2012, Levis2013}.
A {\it charge}, $q$, defined as the number of out-going minus the number of in-coming arrows, 
can be attributed to each single vertex configuration~\cite{Ryzhkin2005,Castelnovo2008}.
Accordingly, in the six-vertex model vertices have zero charge and all other vertices breaking the ice-rule have a net (positive or negative)
charge.

The energy of each vertex configuration is quantified by the Hamiltonian 
$ H=\sum_{k=1}^{16}\epsilon_{k}n_{k}$,
where $n_{k}$ is the number of vertices of type $k$ and $\epsilon_k$  its energy. 
We assign a (not normalized) Boltzmann weight $\omega_k = e^{-\beta\epsilon_{k}}$ to each of the $k=1,\dots,2^{4}$
four-arrow vertex configurations (note that $\omega_k$ can be greater than one if $\epsilon_k$ is negative). We set
$\omega_1=\omega_2=a$, $\omega_3=\omega_4=b$, $\omega_5= \omega_6=c$
for the ice-rule vertices and $\omega_7=\omega_8=d$,
$\omega_9=...=\omega_{16}=e$ for the 2-fold and 1-fold defects,
respectively, ensuring invariance under reversal of all arrows 
(see Fig.~\ref{vertex_configurations}). 

The static phase diagram of this model was obtained in~\cite{Foini2012} by using the cavity analytic method (Bethe-Peierls approximation)
and numerical simulations on $2D$. We will not repeat all the results found in this 
(long) paper but we simply recall that for $d\neq 0$ and $e\neq 0$ the model has a complex phase diagram with many phases.
In particular, we singled out a conventional disordered paramagnetic (PM)
phase, two ferromagnetic (FM) phases ($a$ and $b$ dominated, respectively), and two antiferromagnetic (AF) phases 
($c$ and $d$ dominated, respectively). All transitions are of second order if $d\neq 0$ and $e\neq 0$.
Using numerical simulations, FM order dominated by $a$ vertices was found for $a\gtrsim b+c+d+3e$, close to the critical line predicted by the 
analytic calculations. Using the same arguments, AF order dominated by $c$ vertices was found for  $c\gtrsim a+b+d+3e$ and the PM phase 
was found for $a,c,d,e\lesssim\frac{1}{2}(a+b+c+d+3e)$ (for the precise prediction, that we give here only in approximate form, see~\cite{Levis2012}).
For small defect weight, $e,d \ll a,b,c$ the disordered phase is very close to the {\it critical} spin-liquid phase of the six-vertex model.
The closeness to criticality will play an important role in quenches into the disordered phase.
Henceforth we measure the weights in units of $c$: $c\rightarrow 1$ and $a,b,d,e\rightarrow a/c,b/c,d/c,e/c$.

\subsection{Stochastic dynamics}

We mimic the effect of thermal fluctuations in spin-ice samples by coupling the model to an environment and 
allowing for {\it local single spin flips}  determined by the heat-bath rule. Local moves that  break the spin-ice rule 
are not forbidden and we therefore allow for thermally-activated creation of
defects. The dynamics do not conserve any of the various order parameters, and are ergodic 
for both fixed and periodic boundary conditions. 
We establish a Monte Carlo algorithm
and we define the unit of time as a Monte Carlo sweep (MCs). 
In systems with frustration, as the one we are dealing with,
computer time is wasted by the large rejection of blindly proposed updates. In order to make the computer time 
dynamics faster we use a rejection-free continuous-time Monte Carlo (MC) algorithm~\cite{BKL}.
The longest time that we reached with this method, once translated in terms of usual MC sweeps, is of the order of
$10^{25}$~MCs, a scale that is unreachable with usual Metropolis algorithms. 

Other kinds of dynamic rules have been used in the literature, with different purposes.
For instance, a  rule that preserves the ice constraint does not create defects. In order to sample the
whole phase space on a 
system with PBC in this way one needs to introduce loop updates of any size and winding number.
Such a dynamics have been studied in the 
 3-colouring model on the hexagonal lattice and leads to glassy behavior~\cite{Chakraborty2002, Cepas2012}. Another possible local dynamics which 
 preserve the ice rules would be to 
 update the system by small loops made by four spins around a square plaquette. These dynamics are not ergodic for PBC but they are
 for the six-vertex model with domain-wall BC (DWBC)~\cite{Korepin-ZinnJustin,Zinn-Justin}. For the spin-ice problem, these 
 two possible dynamical models seem quite artificial and do 
 not allow us to study defects' motion in the way that it is observed to occur in the laboratory.   We therefore attach to the moves described in the 
 previous paragraph.
 
\begin{figure}[h]
\centering
\includegraphics[scale=1.2]{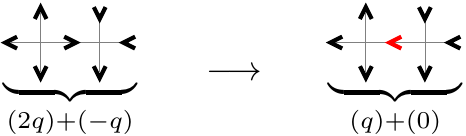} 
\caption{(Colour online.) The reaction in eq.~(\ref{eq:AnnDoubleDef2}).}
\label{fig:Reaction2}
\end{figure}

In terms of a reaction-diffusion model the relevant processes taking place during the time-evolution  are:
\begin{eqnarray}
&& (2q)+(-2q)\to(q)+(-q),  \; \Delta E\propto2k_B T \ln(d/e)  ,  \;\;\; \;\;\;\; \label{eq:AnnDoubleDef1} 
 \\
&& (2q)+(-q)  \to (q)+(0),  \;\;\;\;\; \; \Delta E \propto k_B T \ln (d/{a}) ,  \;\;\;\;\;\;\; \label{eq:AnnDoubleDef2}
 \\
& & (q)+(q)  \to (2q) +(0),  \;\;\;\; \;\;\;\;\; \Delta E \propto k_B T \ln (e^2/bd) , \;\; \label{eq:DoubleDef}
  \\
&& (q)+(-q) \to (0)+(0),  \;\;\;\;\;\;\;\; \Delta E \propto k_B T \ {\ln (e^2/ab)} , \;\; \label{eq:AnnDefects} 
\end{eqnarray}
where the energetic change $\Delta E$ associated with each reaction is shown. As $\Delta E=E_F-E_I$, with $E_F=k_BT\ln \omega_F^{-1}$ 
the energy of the final configuration
and $E_I=k_BT \ln \omega_I^{-1}$ the energy of the initial configuration,  one has $\Delta E=k_BT \ln (\omega_I/\omega_F)$.  
{In our simulations we will typically use $a, b \gg e\geq d$ as this choice is more relevant experimentally.}
In the first case, eq.~(\ref{eq:AnnDoubleDef1}), two defects of type 7 and 8 meet to produce two singly (and oppositely)
charged defects with an energetic gain, if $e > d$, which depends on the ratio $e/d$. 
The total density of defects remains constant after this reaction although their type changes.  
An example of the second case, eq.~(\ref{eq:AnnDoubleDef2}),  is shown in Fig.~\ref{fig:Reaction2}:  
a defect of type 7 (charge $q=2$) meets one of type 14 (charge $q=-1$) 
to produce a defect of type 10 (charge $q=1$) and a spin-ice vertex of type 2 with no charge. 
This corresponds to an energetic gain $\Delta E<0$ which depends  on  {$d/a$}. 
Note that the number of single charged defects has not been modified during this process but the 
number of doubly charged defects diminished and so did the total number of defects.
The reaction in the third line represents an initial state made of the 13th vertex (on the left) 
and the 15th vertex (on the right) that turn into a state with the 3rd vertex (on the left) and the 7th vertex
(on the right) by reversing the internal link. The energy variation is then $\Delta E=k_BT \ln (e^2/bd)$, and this 
can be positive or negative depending on $e/b$ (smaller than one) {\it vs.} $e/d$ (larger than one).  
The fourth reaction is realised, for example, by the reversal of the spin on the edge linking vertex 
number 13 (on the left) with vertex number 14 (on the right), leading to vertex number 3 (on the left)
and vertex number 2 (on the right). The energy change is then $\Delta E = k_BT \ln (e^2/ab)$, a negative quantity for 
our choice of parameters.

In conclusion, with our choice of parameters, 
the reactions in eqs.~(\ref{eq:AnnDoubleDef1}), (\ref{eq:AnnDoubleDef2}) and (\ref{eq:AnnDefects}) 
lead to a decrease in energy while the reaction in eq.~(\ref{eq:DoubleDef})  may be energetically 
favourable or not depending on the ratio $e^2/bd$.

\subsection{Quench dynamics}

We will analyse the  system's evolution after an infinitely rapid quench from a disordered initial 
condition into the disordered (D), $a$-ferromagnetic ($a$-FM) and $c$-antiferromagnetic ($c$-AF) phases. In practice, we choose
a completely disordered configuration ($T_0\to\infty$, $a=b=d=e=1$) as an initial condition; such a state 
is constructed by placing arrows at random on each edge of the square 
lattice $\mathcal{V}$.  If we impose PBC, the number of positive and negative charges is identical. We subsequently 
evolve the MC code with parameters that belong to the three interesting phases.
The system remains globally neutral during the evolution, since  it  is updated by single spin flips which cannot create 
any excess of charge.  

After a quench into the disordered phase the system could be expected to equilibrate relatively rapidly; 
still, it was shown in~\cite{Levis2012} that it remains blocked in metastable states
with a finite density of defects for long times, if the weight of defects is low enough. 
We will investigate this problem in depth here. We will 
demonstrate that metastability also exists after quenches into the ordered phases. Eventually, the
interactions between the spins, mediated by the choice of vertex
weights, creates ordered domains of FM or AF kind.  The quantitative 
characterisation of growth in the ordering processes is given by  two 
possibly different growing lengths extracted from correlation functions 
along orthogonal directions $\parallel$ and $\perp$ that we identify in 
Fig.~\ref{fig:CorrLattice}. 

\subsection{Observables}

The relaxation dynamics of clean lattice systems are usually studied in terms of time-dependent macroscopic observables averaged over different 
realisations of the dynamics (thermal noise, initial conditions) denoted by $\langle ... \rangle$. In particular, we compute the following quantities:
\vspace{0.3cm}
\\
(i) The \emph{density of vertices} of each type: 
\begin{equation}
 n_a(t) =\langle n_1(t)+n_2(t) \rangle \ , \\
 n_b(t) =\langle n_3(t)+n_4(t) \rangle \ ,
 \end{equation}
 \begin{equation}
 n_c(t)=\langle n_5(t)+n_6(t) \rangle \ , 
 \end{equation}
 \begin{equation}
n_d(t) =\langle n_7(t)+n_8(t) \rangle \ , \  \\
 n_e(t) =\langle \sum_{k=9}^{16}n_k(t) \rangle \ ,
 \end{equation}
 \begin{equation}
 n_{def}(t)=n_d(t)+n_e(t) 
\ .
\end{equation}
\\
(ii) The \emph{two-times self-correlation function} defined by:
\begin{equation}
C(t,t_w)=\frac{1}{2L^2}\sum_{(i,j)\in\hat{\mathcal{V}}}\langle S_{(i,j)}(t)S_{(i,j)}(t_w) \rangle   
\end{equation}
with $t>t_w$. The indices $(i,j)$ denote the coordinates of an Ising spin in the medial lattice $\hat{\mathcal{V}}$ 
(i.e. the vertices of the square lattice shown in red in Fig.~\ref{fig:CorrLattice}). 
\vspace{0.3cm}
\\
(iii) The {\it space-time correlation functions}. The definition of the relevant correlation functions between 
different points in the lattice is not straightforward when we introduce some anisotropy in the model (for example,
 by choosing $a>b$). For convenience, we define a set of correlation functions between spins using two different 
 orientations: along the Cartesian axes $\vec{u}_x$ and $\vec{u}_y$ and along the $\pi/4$-rotated axes 
 $\vec{u}_{\parallel}$ and $\vec{u}_{\perp}$ (see Fig.~\ref{fig:CorrLattice}).
The space-time self correlation functions along $\vec{u}_{\parallel}$ and $\vec{u}_{\perp}$ are defined as
\begin{equation}
 G^{\parallel}(r=n,t)=\frac{1}{L^2}\sum_{(i,j)\in\hat{\mathcal{V}}}\langle S_{(i,j)}S_{(i,j+n)} \rangle \ ,
 \end{equation}
 \begin{equation}
 G^{\perp}(r=n,t)=\frac{1}{L^2} \sum_{(i,j)\in\hat{\mathcal{V}}} \langle S_{(i,j)}S_{(i+n,j)} \rangle \ , 
\end{equation}
where $n \in \mathbb{N}$. 
\vspace{0.3cm}
\\
(iv) The \emph{growing lengths} $L^{\parallel}(t)$ and $L^{\perp}(t)$ along $\vec{u}_{\parallel}$ and $\vec{u}_{\perp}$ 
are extracted numerically from the scaling of the space-time correlations: 
\begin{equation}
G^{\parallel,\perp}(r,t)\simeq F_{G^{\parallel,\perp}}\left(\frac{r}{L_{\parallel,\perp}(t)}\right) \ .
\end{equation}

\section{Defect density}\label{sec:Metastability}

In this Section we study the density of defects, vertices of type $e$ and $d$ after quenches into the different 
phases. With this analysis we investigate the possibility of finding long-lived metastable states after dynamic quenches from a
fully disordered initial condition.

\subsection{Quench into the PM phase}\label{sec:QuenchPM}

In the following, we study the evolution of the model after a quench from a random initial condition ($a=b=d=e=1$) into a different PM state, 
typically close to the SL critical phase 
(i.e. $a=b=1$ and $d,e \ll 1$). In the initial configurations defects are common: we are interested here in the mechanisms leading to their annihilation. 

\subsubsection{Equally probable defects, $d=e$.}

The evolution of the system after a sudden quench into the PM phase with $d=e$ was already reported in~\cite{Levis2012}. 
Let us recall some useful results:  
for values of $d$ large enough ($d=e\geq 10^{-4}$) the density of defects $n_{def}$ quickly saturates to its equilibrium value. For smaller $d$'s, the 
system gets arrested  into a 
metastable state with finite and constant density of defects $n_{def}=n^p$ for long periods of time. This dynamical plateau lasts longer for smaller $d$s, 
a behaviour reminiscent to 
what was found in $3D$ dipolar spin-ice~\cite{Castelnovo2010}. 
The time regime where the density of defects finally leaves the plateau and reaches its equilibrium value,  is characterised by a scaling of the dynamic 
curves with the characteristic time $d^{-2}$.
This scaling strongly suggests that the relevant time scale in the system  is the typical time needed to create a pair of single defects. From an ice-rule 
state, the energy change 
associated with the reaction: $(0)+(0)\to (q)+(-q)$ is, e.g., $\Delta E\propto -k_B T \ln  (d^2/ab)$, with $a=b=1$ in the simulation. 
Then, by a simple Arrhenius argument, 
the typical time needed in order to overcome this barrier is  $\propto \exp(\beta \Delta E)$ giving the before mentioned time-scale $\tau \propto d^{-2}$.   

At a first sight, one could think that the emergence of this dynamical plateau in the density of defects is due to the presence of doubly charged defects, the 
ones with $q=\pm 2$ and weights $\omega_7=\omega_8=d$. Indeed, doubly charged defects 7 and 8 must decay into two single charged defects in 
order to be able to move. However, the inverse
reaction $(2q)+(0) \to (q) + (q)$ in eq.~(\ref{eq:DoubleDef}) is accompanied by an increase in energy, $\Delta E\propto k_BT\ln(b/d)>0$ when $d=e$, and it is 
energetically unfavourable. Therefore, $d$-vertices get naturally stuck in the sample and are very hard to eliminate. This mechanism could give a 
justification for the plateau in the case $d=e$.

In real spin-ice realisations, both in $2D$ and $3D$, the energy associated to doubly charged defects is much larger than the one of 
single charged ones $\epsilon_{7,8} 
\gg \epsilon_{9,\dots,16}$. It is then more relevant to experiments to study in detail the effect of $d<e$ 
in the time evolution of the model,
and to revisit the influence of $e$ and $d$ defects on the development of the plateau. 

\subsubsection{Single charged defects are more favourable than doubly charged ones,  $d< e$.}

We investigate now the dynamical consequences of choosing different weights for the two kinds of defects.
We focus on the fate of the dynamical plateau when doubly charged defects are rapidly suppressed, with $d$ the 
smaller weight in the model. The inspection of the reaction rates for the annihilation-creation of defects suggests to study the two 
following cases separately:
\begin{itemize}
\item $d<e$ and $d>e^2$: Single charged defects $e$ are slightly more favourable than $d$-defects. However, the decay of $d$-defects into two $e$-defects 
following the inverse reaction eq.~(\ref{eq:DoubleDef}) must overcome an energy barrier, as 
$\Delta E=k_BT \ln(bd/e^2)>0$. 
 \item $d<e$ and $d< e^2$:  Doubly charged defects are very unfavourable. The decay of $d$-defects into two $e$-defects is energetically favoured and 
occurs spontaneously. 
\end{itemize}

\begin{figure}[h]
\centering
\includegraphics[scale=0.35,angle=0]{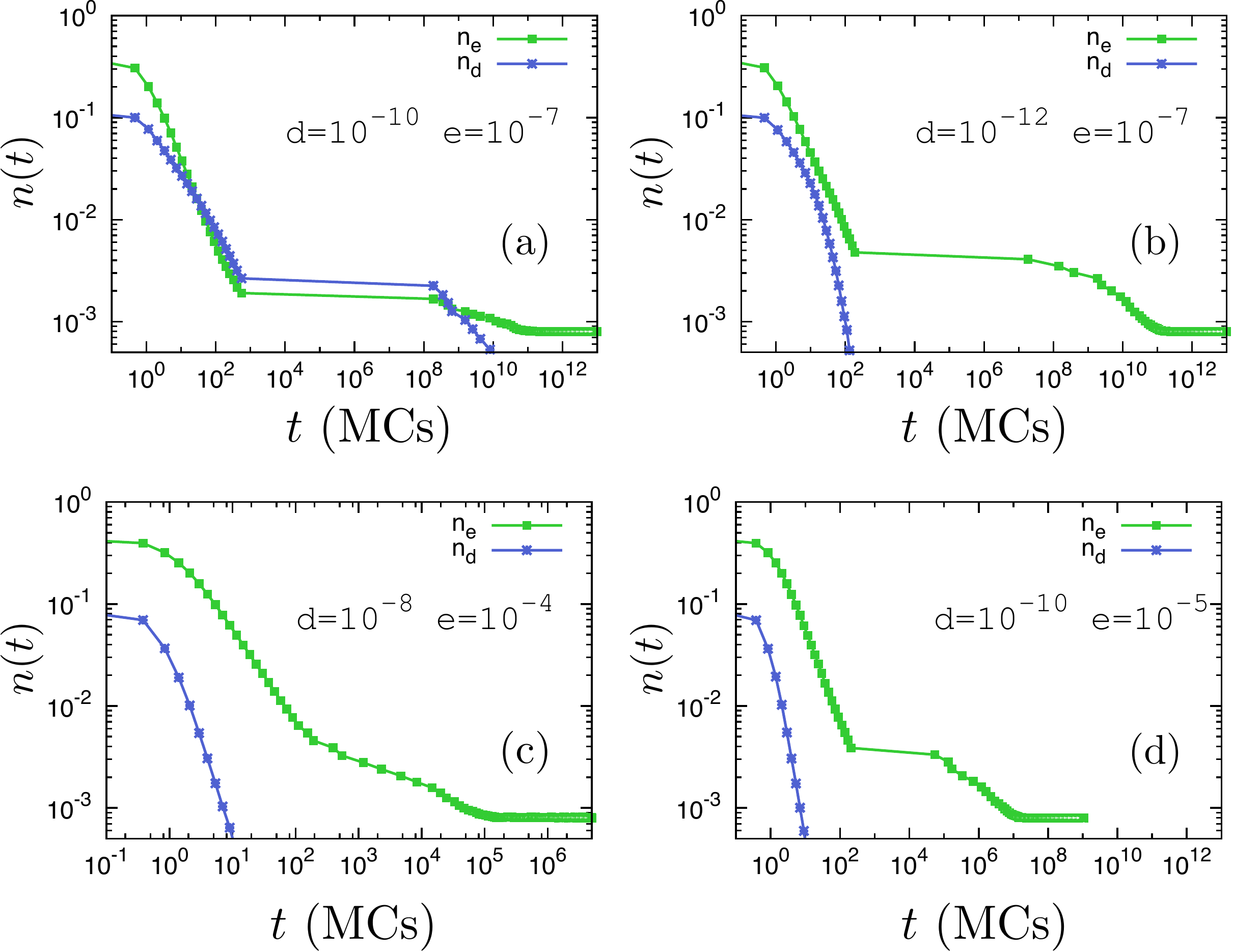} 
\caption{(Colour online.) Quench into the PM phase.
Decay of the density of vertices for $a=b=1$ and $d< e$ for a square lattice with $L=50$ averaged over $500$ realisations. 
The weights of the defects are indicated on the figure. The 
panels in the first row, (a) and (b), are for parameters such that $d>e^2$, with $e^2/d=10^{-4}$ in (a) and $e^2/d=10^{-2}$ in (b). 
The panels in the second row, (c) and (d), are for $d=e^2$.}
\label{fig:deltaE>0}
\end{figure}

As shown in Fig.~\ref{fig:deltaE>0} (a), for $d>e^2$ and large enough values of $d$, $d/e^2 \gtrsim 10^2$,
the decay of $n_e$ (green data points) and $n_d$ (blue data points) freeze at a 
metastable density for around five decades in time. The density of $e$-vertices $n_e$ is smaller than $n_d$ in the plateau regime. 
Instead, still for $d>e^2$  but for smaller values of $d$, $d/e^2 \lesssim 10^2$,  $d$-vertices rapidly disappear and the plateau is 
only seen on
$n_e$, as shown in panel (b). Indeed, after a rapid decay, $n_e$ gets frozen into a metastable value for a long time before it finally 
reaches its equilibrium value. 
Hence, one can conclude that the presence of $d$-defects in the system {\it is not} responsible for the emergence of the dynamical 
plateau
in the total density of defects $n_{def}$.

The data in panels (c) and (d)  in Fig.~\ref{fig:deltaE>0} were obtained for $d=e^2$. 
The density of $n_e$ remains larger than $n_d$ during the whole evolution in both cases. 
Similarly to what was observed for $e=d$~\cite{Levis2012}, the system gets blocked into a metastable plateau only for small enough values of 
$e \lesssim 10^{-4}$, and the existence of this arrested dynamical regime is not due to an excess of $d$-vertices.  
The evolution of $n_e$ and $n_d$ for $d<e^2$ 
shown in Fig.~\ref{fig:deltaE>0} supports this observation. Although $n_d$ rapidly vanishes, $n_e$ exhibits a dynamical arrest. The value of 
the plateau density can, 
in principle, depend on the weight of the vertices in a complicated manner. We did not study this last point in detail here.   

In short, the system can be arrested in long-lived metastable states with many defects of type $e$.

\begin{figure}[h]
\centering{
\includegraphics[scale=0.3]{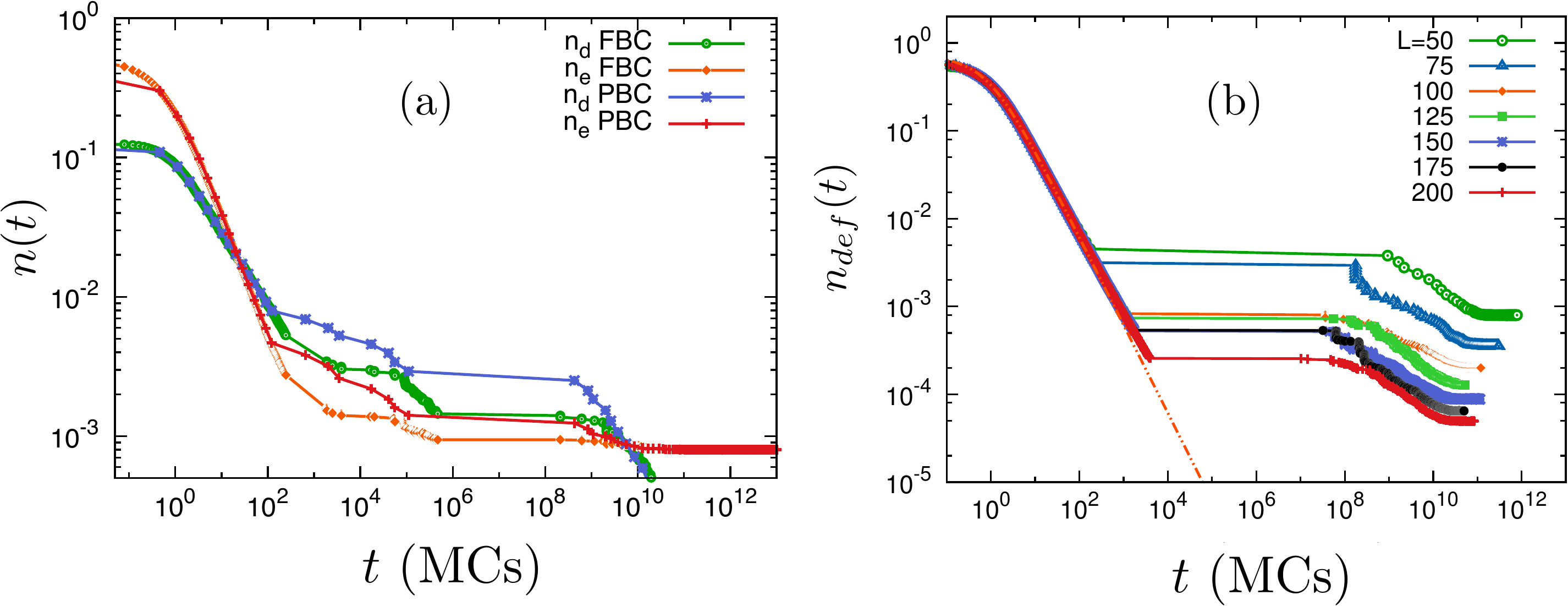}
\caption{(Colour online.) Quench into the PM phase.
(a) Time evolution of the density of defects, $n_d$ and $n_e$, for different boundary conditions: periodic boundary conditions (PBC) and 
fixed boundary conditions (FBC). The data were obtained with an average over $500$ realisations of 
the dynamics with $a=b=1$ and $d=e=10^{-7}$ for a system with linear size $L=50$. 
(b) The time-dependence of the total density of defects, $n_{def}=n_d+n_e$, 
for different system sizes with PBC at $d=e^2=10^{-14}$, 
$a=b=1$. The linear sizes are given in the key and vary by $25$ from $L=50$ to $L=200$. The data were obtained after averaging 
over $300$ runs. 
The initial decay of $n_{def}$ is confronted to 
$\rho_0/(1+\Omega t)$, with $\Omega$ a fitting parameter (red dotted line), see the text for a discussion. 
}
\label{fig:FBCvsPBC}
}
\end{figure}

\subsubsection{Boundary conditions}

In order to better understand the emergence of the frozen regime we repeated the numerical experiment with fixed boundary conditions (FBC): the state of each 
spin on the boundary is kept fixed from the one it had in the (random) initial configuration during the simulation. One has to be careful when choosing the 
boundary conditions and make sure that these do not induce a polarisation of the sample. Indeed, polarised boundary conditions such as the DWBC can have 
dynamical consequences such as the drift of 
defects (loosely speaking, magnetic monopoles). These effects have not been studied here.

In the initial high temperature state, defects of any kind populate the system. After the quench, one of the mechanisms for relaxation is 
the annihilation of oppositely charged defects. In 
order to do so, defects have to meet in the appropriate manner, meaning that the reversal of the spin shared by both of them restores the ice rule. In the reaction-diffusion language 
this corresponds to the process $(q)+(-q)\to (0)+(0)$. Two defects of opposite charge $\pm1$ can also meet in the `wrong' way and create, by a single spin-flip, a pair of doubly 
charged defects accordingly to: $(q)+(-q)\to (2q)+(-2q)$. 
Starting from a completely ordered FM configuration, one can create a pair of defects by flipping a string of spins. The latter can wind around the lattice by PBC. 
Then, in order to annihilate these pair of defects, one must flip back all the spins belonging to the string. One can imagine that this kind of extended structures  could
be responsible for the slowing down of the dynamics. If so, the evolution of the system with FBC, where winding strings are absent, should not present a dynamical plateau. 
As shown in Fig.~\ref{fig:FBCvsPBC} this is not the case: a metastable plateau in the evolution of the density of defects appears with FBC as well. This is due to the fact that, 
in the presence of more than a single pair of defects, there is always a way to annihilate all the defects without going through the boundaries of the lattice. In this sense, 
the dynamics are insensitive to the nature of the boundary conditions. 

\subsubsection{Finite size effects}

As already pointed out in~\cite{Levis2012}, the metastable density of defects for $d=e$ depends on the linear size of the system. One should then ask, for 
generic parameters, whether 
the observed metastable density is just a finite size effect or not. In order to give an answer to this question we simulated systems of different sizes under the same 
conditions that we chose to be $a=b=1$ and $d=e^2=10^{-14}$. The results obtained are shown in Fig.~\ref{fig:FBCvsPBC}~(b). 
The height of the plateau, $n^p$, and the time 
spent by the system in this regime decreases with the size of the system. However, as the plateau height is subject to strong fluctuations, we are not able to 
predict its precise dependence on the system size. The data do not show saturation at a finite value $n^p$ nor length of the plateau and this 
gives a strong indication that this effect is due to the finiteness of the samples.   
This is confirmed by the data obtained after a quench into the $c$-AF phase for different system sizes (see Fig.~\ref{fig:PlateauAF}): 
in this phase the data are less noisy and we found that the plateau density $n^p$ 
depends on the size of the system as $\propto L^{-2}$ which vanishes in the thermodynamic limit.  Having said this, we wish to 
stress that the system sizes of artificial spin-ice samples are of the same order as the ones used in our numerics and, therefore, blocking
effects of the kind here shown are expected to exist in those samples as well, e.g. $L\approx 150$ in the samples studied in~\cite{Wang2006}.
\subsubsection{Initial decay}

For $d=e$, see ~\cite{Levis2012}, the initial decay of $n_{def}$ is well fitted by a power law decay 
\begin{equation}
n_{def}(t)=\frac{\rho_0}{(1+\Omega t)^{\alpha}}
\label{eq:power-law-decay} 
\end{equation}
with $\alpha=0.78$. The evolution of $n_e$ for $d=e^2$ shown in Fig.~\ref{fig:FBCvsPBC}~(b) is rather well fitted by a diffusive decay 
$\rho_0/(1+\Omega t)$. 
Interestingly, when $d\leq e^2$ the decay of the defects' density agrees with the mean-field reaction-diffusion picture, $n_{def}(t) \simeq \rho_0/(1+\Omega t)$,
proposed in~\cite{Castelnovo2010} for $3D$ dipolar spin-ice. The presence of $d$-defects modifies this behaviour and makes the decay 
slower~\cite{Levis2012}, i.e. 
$\alpha$ decreases. This suggests that the exponent $\alpha$ depends on $d/e^2$ and crosses over from $\alpha=1$ for $d/e^2\ll 1$ 
($d\leq e^2$) to $\alpha<1$ beyond 
this limit.

\subsection{Quench into the FM phase}\label{sec:QuenchFM}

\begin{figure}[h]
\centering
\includegraphics[scale=.3]{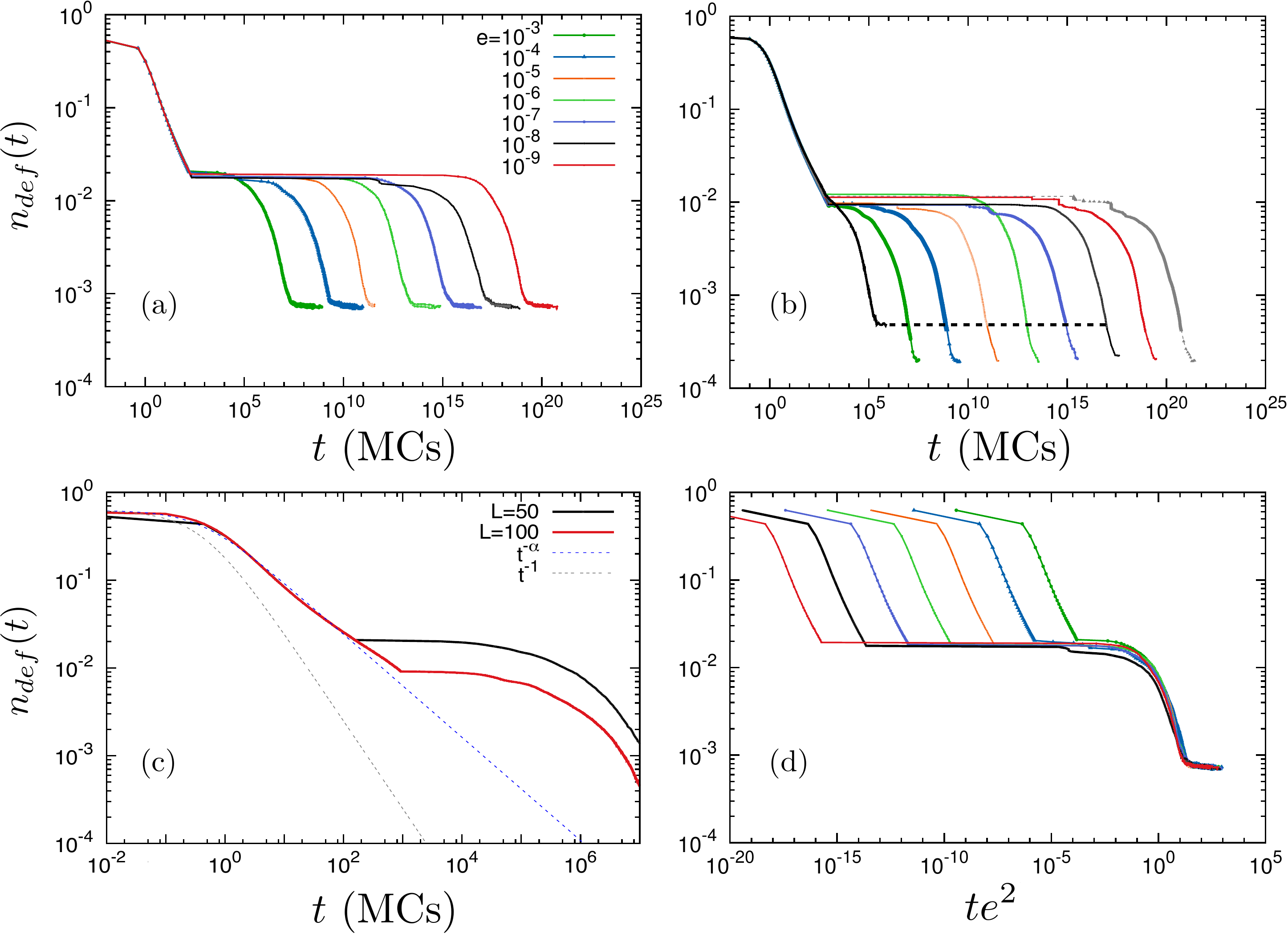} 
\caption{(Colour online.) Quench into the FM phase. Time dependent density of defects after a quench from $a=b=d=e=1$ to $a=5$, $b=1$ and $d=e^2$ for the different values of $e$. 
(a) $L=50$ and $e=10^{-3},10^{-4}, ..., 10^{-9}$ (as shown in the key). (b) $L=100$ for $e=10^{-2}$ (in black)$, 10^{-3}, ..., 10^{-10}$. 
For $e=10^{-2}$ the system saturates to its equilibrium value shown with a dotted black line. (c) Short-time behaviour for $e=10^{-3}$,
 $L=50$ and $L=100$. The decay is confronted to  $\rho_0/(1+\Omega t)$ (grey dashed line) and the fit $\rho_0/(1+\Omega t)^{\alpha}$ 
 with $\alpha=0.59$ (blue dashed line).  (d) Test of scaling with $t e^2$ for $L=50$.}

\label{fig:PlateauFM}
\end{figure}

We now turn to the ordering dynamics following a quench from a random initial condition into the FM phase dominated by 
$a$-vertices (i.e. $a\gtrsim b+1+d+3e$ as discussed in detail in~\cite{Foini2012}). 
As shown in Fig.~\ref{fig:PlateauFM}, a dynamic arrest occurs for small defects' 
weights during the relaxation towards the $a$-FM phase as well. We anticipate that  the same kind of behaviour also appears 
when the system is  quenched into the $c$-AF phase (see Fig.~\ref{fig:PlateauAF}).    


In this section we analyse the relaxation towards the FM phase by studying the decay of the defects' density for different values of the external 
parameters. In Fig.~\ref{fig:PlateauFM} we show the evolution of the density of defects after a quench to $a=5$, $b=1$, $d=e^2$ and different values of $e$ for two different 
system's sizes $L=50$ (a) and $L=100$ (b). The data shown have been averaged over $10^3$ independent realisations of the dynamics. 
For small enough $e$ ($e\lesssim 10^{-3}$) the system gets frozen into a metastable state with a finite density of defects. Similarly to what is observed after the quench into the  PM 
phase~\cite{Levis2012}, the time the system spends in this plateau is longer for smaller $e$. 
The ordering process following a quench into the FM phase is characterised by a time scale $\tau\propto e^{-2}$ in the regime in which 
$n_{def}$ leaves the plateau. As shown in 
Fig.~\ref{fig:PlateauFM} (d) for $L=50$ all the curves collapse into a single curve when rescaling the time variable by $\tau$. Therefore, the typical time associated with the creation 
of a pair of defects is the relevant time scale during this time regime. 

\begin{figure}[h]
\centering
\includegraphics[scale=.35]{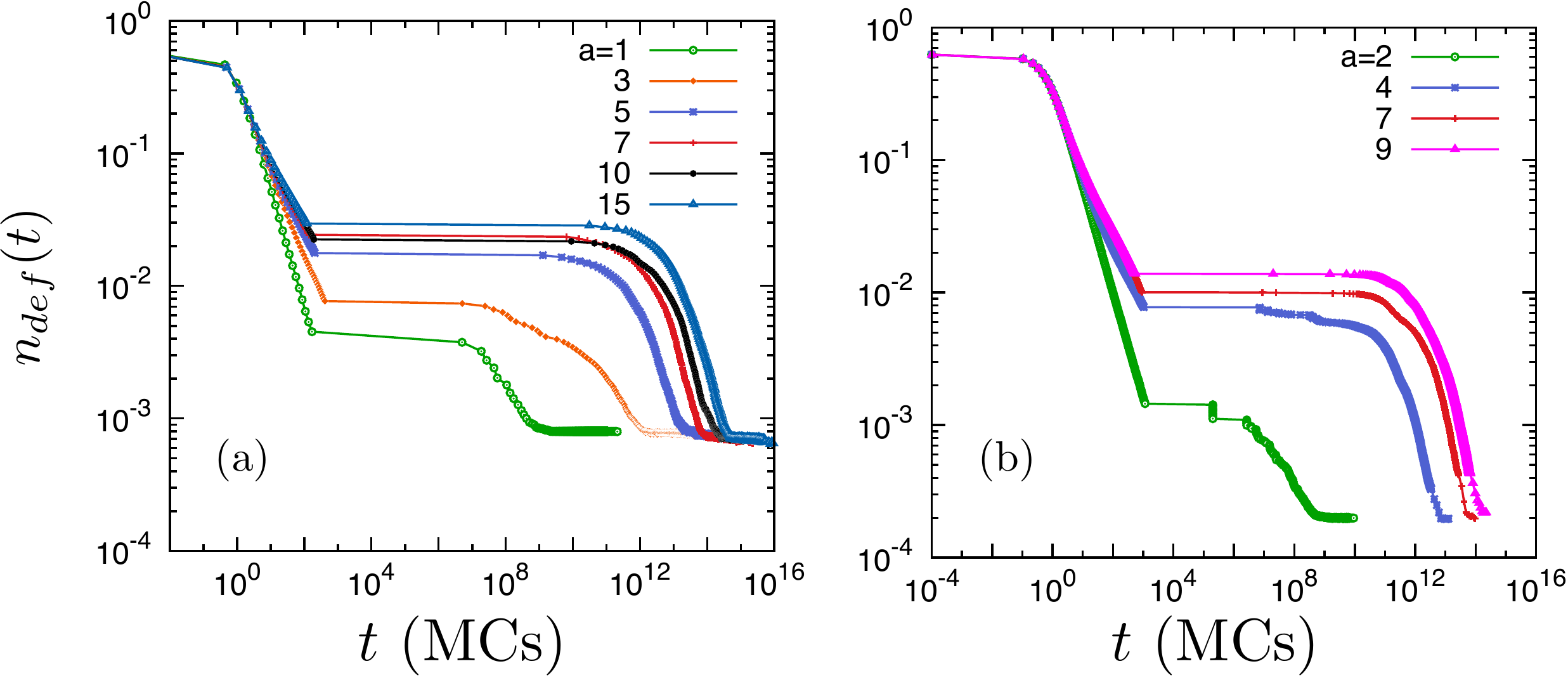} 
\caption{(Colour online.) Quench into the FM phase. Time-dependent density of defects after a quench from $a=b=d=e=1$ to $b=1$, $d=10^{-18}$, $e=10^{-6}$ and different values of $a$. 
(a) $L=50$ for $a=1,3,5,7,10,15$ (as shown in the key). (b) $L=100$ for $a=2,4,7,9$. The data have been averaged over $300$ independent runs.}
 \label{fig:PlateauFM_as}
\end{figure}

The evolution of the density of defects following a quench into different points of the $a$-FM phase is shown in 
Fig.~\ref{fig:PlateauFM_as} for $L=50$ (a) and $L=100$ (b) samples.  
During a short time regime ($t \lesssim 10$ MCs) the density of defects decays
independently  of $a$. 
For later times, the decay  depends on the value of $a$. In particular, the expected power-law decay $n(t)\sim t^{-\alpha}$ becomes slower for larger values of $a$. Therefore, the 
exponent $\alpha$ depends on the weights of the vertices and decreases when increasing $a$. 
The metastable density of defects increases with $a$ and depends on the system size. 

The initial decay of $n_{def}$ can be fitted by the power-law in eq.~(\ref{eq:power-law-decay}) with $\alpha\simeq 0.59$ over the whole time regime 
before the system reaches the plateau density [as shown in Fig.~\ref{fig:PlateauFM}~(c)]. In the $a$-FM the decay  becomes slower than the diffusive law, $t^{-1}$, found in the 
disordered phase. The algebraic decay does not depend on the size of the system as  suggested by the data shown in Fig.~\ref{fig:PlateauFM}.

A general statement can be made at this point: the value of the exponent $\alpha$ decreases - and hence the relaxation becomes slower - when 
going deeper into an ordered phase (of FM or AF kind). 

\subsection{Quench into the $c$-AF phase}\label{sec:QuenchAF}

\begin{figure}[h]
\vspace{0.4cm}
\centering
\includegraphics[scale=.38, angle=0]{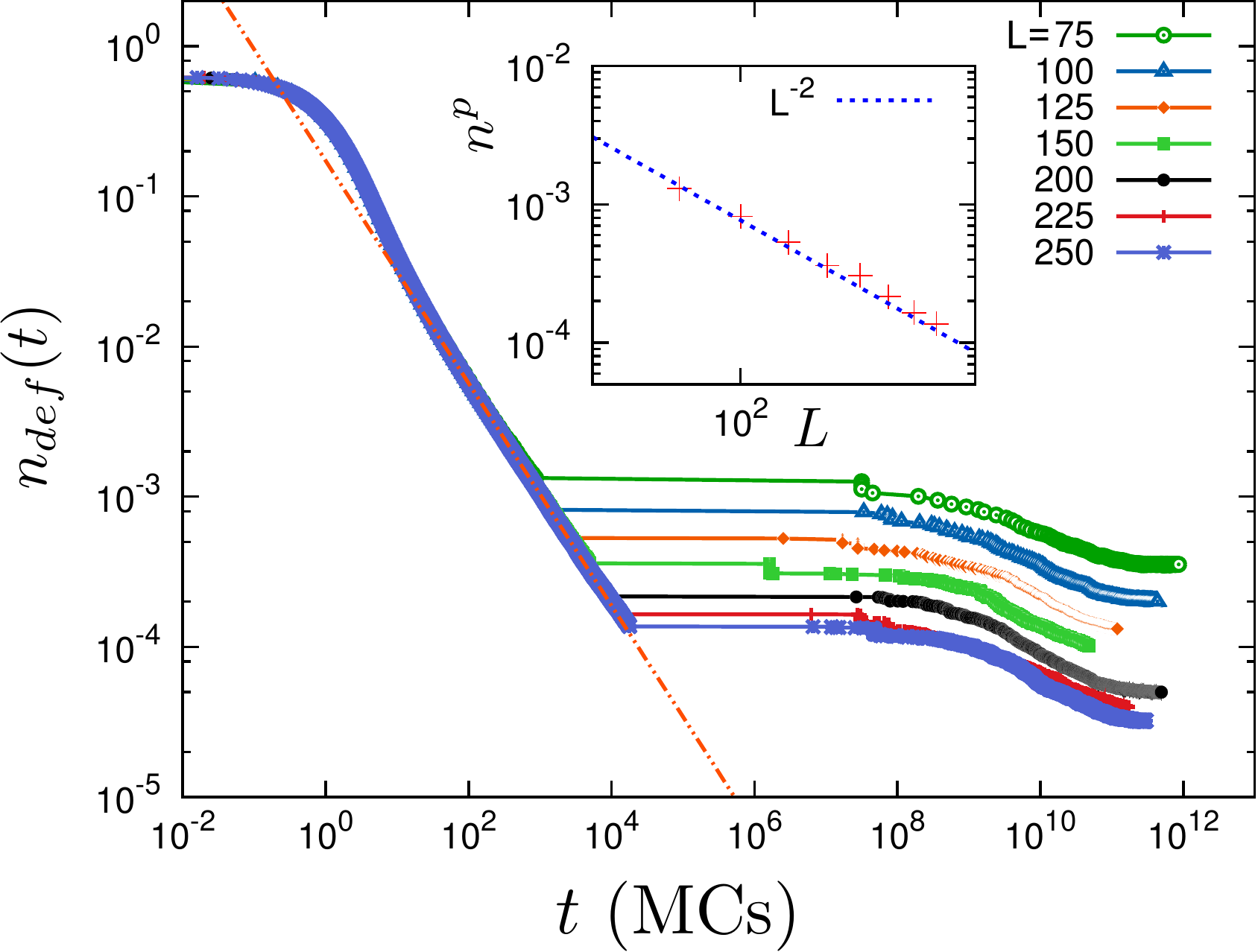} 
\caption{(Colour online.online.)) Quench into the $c$-AF phase.
Temporal decay of the density of defects in the system following a quench to the parameters $d=e^2=10^{-14}$, $a=b=0.1$ and $c=1$.
The data plotted were obtained by using different lattice sizes $L=50,...,250$ and averaging over $300$ independent realisations of the dynamics. The initial 
decay is confronted to the power-law decay $t^{-\alpha}$ with $\alpha=0.74$.}
\label{fig:PlateauAF}
\end{figure}

We follow now the evolution of the system after a quench from a random initial condition into the AF phase dominated by $c$-vertices 
(i.e. $1\gtrsim a+b+d+3e$~\cite{Foini2012}). 
Figure~\ref{fig:PlateauAF} displays the temporal dependence of the total density of defects, $n_{def}$, after such a quench for systems 
with different linear sizes given in the key. The inset shows the linear size dependence of the plateau height extracted from the data
in the main part of the figure in a double logarithmic scale. The data points are accurately fitted by a $L^{-2}$ dependence that 
suggests that the plateau will disappear in the thermodynamic limit. The initial decay is algebraic with a non-trivial power $t^{-\alpha}$ with 
$\alpha=0.74$ independently of the system size.

\subsection{Conclusion}

In this section, we have shown that the system gets arrested into a long-lived frozen state for all kind of quenches as long as  the weights $e$ and $d$ are 
small enough ($d\leqslant e\lesssim10^{-4}$).  After inspection of the persistence of the plateau for a large range of parameters and different boundary 
conditions, we 
concluded that the emergence of such dynamical arrest is not due to the presence of doubly charged defects and it is not an artifact of the periodic boundary 
conditions. The metastable density depends on the system size, in such a way that it might disappear in the thermodynamic 
limit. This assumption is  supported by the $L^{-2}$ dependence of the metastable density of defects after a quench into the $c$-AF phase 
(see Fig.~\ref{fig:PlateauAF}).  The scaling $n_{def}(t)\sim f(te^2)$ shown in Fig.~\ref{fig:PlateauFM} (d) gives a simple interpretation of 
the long time dynamics: the creation of a pair of defects is needed to `unblock' the evolution and allows the system to reach its equilibrium state. 
Another result reported in this section concerns the initial decay of defects right after the 
quenches. We have shown that the density of defects follows a power-law decay characterized by an exponent $\alpha$ which depends on the vertex 
weights.  In the PM phase, one recovers the mean field $t^{-1}$ decay for $d\leqslant e^2$, as reported in 3D dipolar 
spin-ice~\cite{Castelnovo2010}. For 
$d=e$ a smaller exponent $\alpha= 0.78$ was found~\cite{Levis2012}, indicating that the relaxation in qualitatively slower. After a quench into the FM and 
AF phases 
the decay of defects becomes slower when choosing parameters deeper into the ordered regions of the phase diagram. For $a=5$, $b=1$ and 
$d=e^2=10^{-6}$ 
we found $\alpha=0.59$ and we further showed that, at fixed value of $b,d,e$, this exponent decreases when increasing $a$.  

\section{Coarsening dynamics and ageing}\label{sec:Coarsening}

In this Section we analyse the ordering process following a quench into the
FM and AF phases from a totally random initial condition. We choose to work with defect weights satisfying
$d=e^2$, differently from what we presented in~\cite{Levis2012}, where $d=e$. The interest is to investigate the role payed by each
kind of defect in the ordering dynamics, for parameter that are closer to the experimental ones in ASI~\cite{Morgan2011,Morgan2013}.

\subsection{Slow relaxation towards the PM phase}
\label{subsec:slow-relax}

\begin{figure}[h]
\centering
\includegraphics[scale=0.285,angle=0]{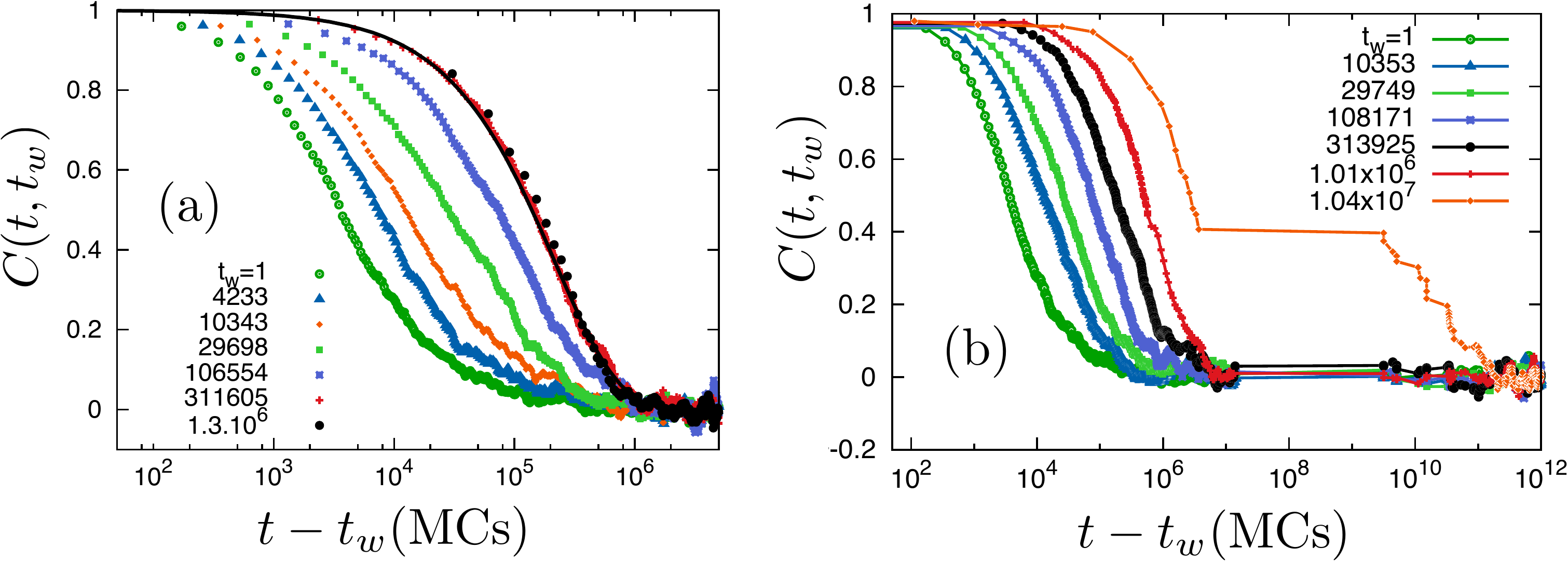}
\caption{(Colour online.) 
Quench into the FM phase from a random initial configuration. 
Two-time self-correlation function $C(t,t_w)$ for a system 
with $L=50$ and data averaged over $500$ realizations. 
The system evolves with $a=b=1$ and $d=e^2 = 10^{-4}$ in (a) while $d=e^2 = 10^{-12}$ in 
(b). Note that in both cases $d=e^2$. The equilibrium correlation function in (a) is confronted to $\exp[-(t-t_w)^{\gamma}/\tau_e]$ with $\gamma=0.81$ and 
$\tau_e=23014$ (solid black line), a law that describes the data very accurately.}
\label{fig:CTT}
\end{figure}

In Fig.~\ref{fig:CTT}  we show the decay of the two-time correlation function $C(t,t_w)$ 
as a function of the time difference $t-t_w$ for different values of $t_w$ shown in the key,
and  working parameters such that $a=b=1$ and
$d=e^2$ (with different choices of the defect weights in the two panels, $d=e^2=10^{-4}$ in (a) and $d=e^2=10^{-12}$ in (b)).
Recalling the results shown in Fig.~\ref{fig:deltaE>0}~($\mbox{c}$) and (d), metastability (a plateau in the number of defects) 
is not expected in (a) as the value of $d=e^2$ is sufficiently large, while it is expected after times, say, of the order of 
$10^4$~MCs in~(b) since $d=e^2$ it is pretty small {($e\leq10^{-4}$)}.

One can distinguish different dynamical regimes from these curves. For short times, as long as neighbouring defects 
annihilate in a few MCs the correlations are time translational invariant and close to one. 
At later times time-translational invariance is lost and  the system exhibits non-stationary relaxation. The longer the waiting-time $t_w$ is, 
the slower the decay, as in an {\it ageing} situation.

 In panel (a), when $d=e^2$ is relatively large, 
a stationary regime is attained for waiting-times of the order of $t_w\simeq 3\ \times \ 10^{5}$ MCs as demonstrated by the 
fact that the two curves corresponding to the longest $t_w$, beyond this time-scale, fall on top of each other and do not 
depend on $t_w$. As shown in the figure, the equilibrium curve follows a stretched exponential decay: $C_{eq}(t,t_w)=F(t-t_w)=
\exp[-(t-t_w)^{\gamma}/\tau_e]$ with $\gamma=0.81$ and $\tau_e \approx 25 \times 10^3$ MCs. One could expect a 
similar picture away from the `spin ice' ($a=b=1$) curve $d=e^2$ in the disordered phase, with some dependence 
of the parameters $t_{eq}$, the relaxation time, $\gamma$, the stretching exponent and $\tau_e$, the characteristic time.
We have not studied these dependencies in detail. 

The behaviour is different in panel~(b), where the weight of defects is much smaller. The curves do not 
reach a stationary regime for the waiting times used. Moreover, a long plateau is seen in the curve for 
$t_w\approx 10^7$ MCs, reminiscent of the plateau in the decay of $n_e$ 
shown in Fig.~\ref{fig:deltaE>0}~(d). The system does not evolve during a period of time in between $\approx 10^7$ and $10^9$ MCs.
In this case an extremely long time-decay is necessary to reach a complete decorrelation.
A steady state regime is not reached in the simulation, the correlations continue to evolve
for all waiting times shown.

As argued in Sec.~\ref{sec:Metastability}, the fire sinite of the lattice has very strong effects in these systems. For 
larger system sizes the  cross-over to metastability will be pushed to longer times, and to infinity in the thermodynamic 
limit. The ever-lasting non-stationary relaxation is due to the proximity to the spin-liquid critical phase of the six vertex model when the 
defect weights are small enough. In any critical relaxation the system in question will  grow equilibrium patches with a time-dependent 
critical length. This length will need an infinite time to reach the size of the system if this diverged. 

\subsection{Anisotropic FM domain growth}
\label{subsec:coarsening-FM}

Starting from a random initial configuration, we quench the system by setting $a=5$, $b=1$ and $d=e^2=10^{-10}$ at $t=0$. 
The equilibrium state corresponding to this set of parameters is deep into the FM phase. This choice strongly favours $a$-vertices. 
The system will then evolve towards its ordered FM state by growing domains made of type-$1$ and type-$2$ vertices.

\begin{figure}[h]
\centering
 \includegraphics[scale=0.38]{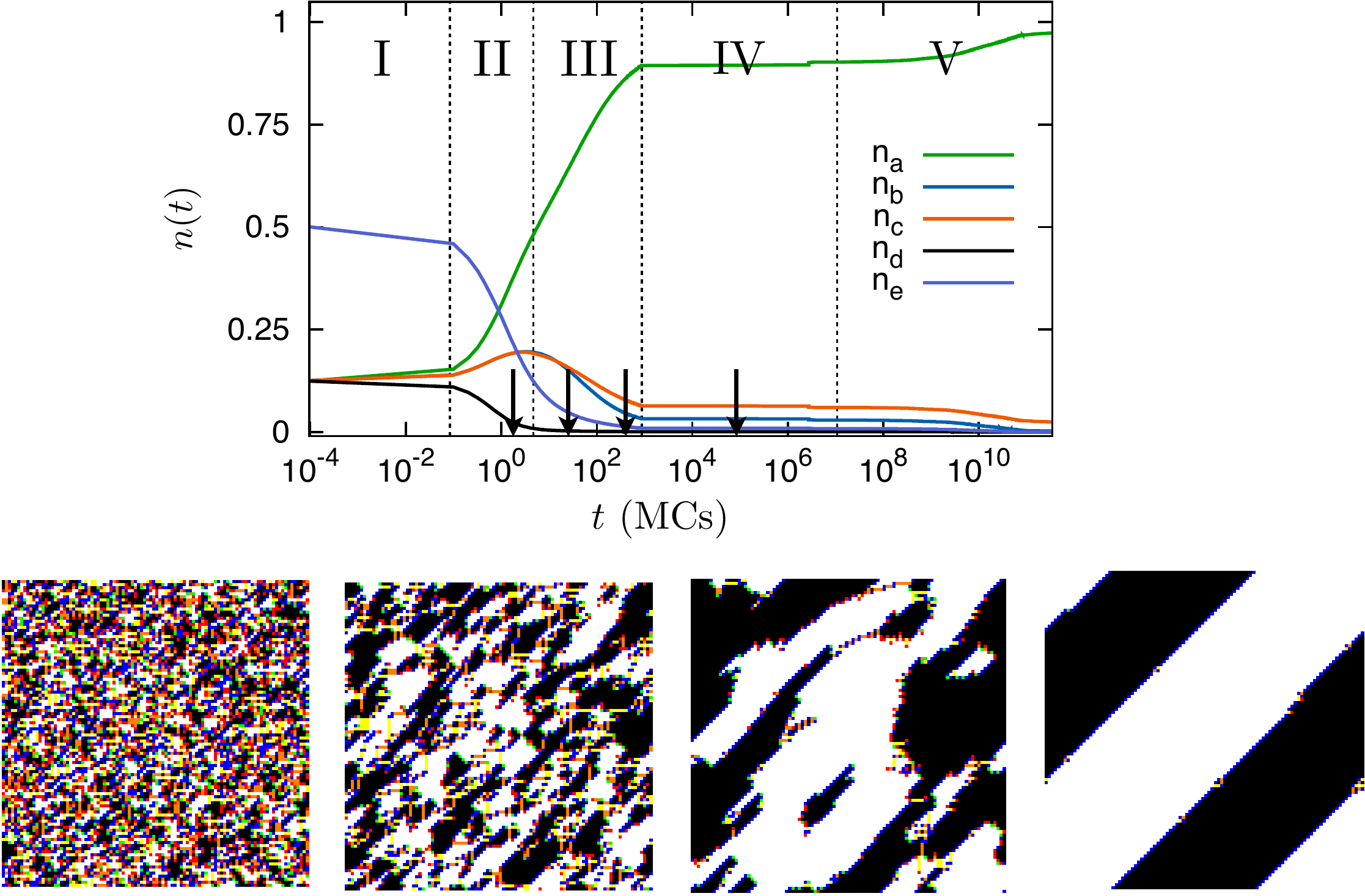}
\caption{(Colour online.) 
  FM ordering.  Upper panel: time evolution of
  the density of vertices with weight $a,b,c,d$, and $e$ for $a=5$, $b=1$,
  $d=e^2=10^{-10}$, in a system with linear size $L=100$. The data are averaged over 300 samples.  
  The snapshots are typical configurations
  at the instants indicated by the arrows. Black and white points are vertices 1 and 2 as defined by the colour code shown 
  in Fig.~\ref{vertex_configurations}.}
\label{fig:Evolution}
\end{figure}

A first hint into the dynamics of the system is given by the time evolution 
of the density of vertices, $n_\kappa(t)$, for each kind of vertex, $\kappa=a,b,c,d,e$. This is 
shown in Fig.~\ref{fig:Evolution}. Differently from what we presented in~\cite{Levis2012} we distinguish here the 
density of each kind of defect as their weights are not the same. 
The data are accompanied by four configurations that illustrate the evolution of the system. 
One clearly observes the growth of anisotropic ordered FM  domains. The directions $\vec u_{\parallel}$ and 
$\vec u_\perp$ defined in Fig.~\ref{fig:CorrLattice} are parallel and orthogonal to the longer domain walls, respectively.  

From inspection of the data plotted in Fig.~\ref{fig:Evolution} we can identify four different dynamical regimes: 
\\
(I) A short time regime ($t \lesssim 10^{-1}$ MCs) during which all densities vary very little
(see, e.g., the data in~\cite{Levis2012} where we gave more data-points on this very short regime, for a different set of parameters).\\
(II) An intermediate regime ($t \lesssim 10$ MCs) characterised by the annihilation of  a large number of defects which are transformed into
ice-rule vertices by a few single spin-flips. Then $n_d$ and $n_e$ decay (independently  
of the value of $a$, as shown in Fig.~\ref{fig:PlateauFM_as})
while $n_a$, $n_b$ and $n_c$ increase. Quite surprisingly $n_b$ and $n_c$ increase in the same way in this 
regime.
The typical configuration shows no apparent order
although the tendency to align in a diagonal direction is already visible. 
\\
(III) A slow relaxation regime in which the dominant dynamical mechanism is the one of growing anisotropic FM domains made by type-1 (black) 
or type-2 (white) vertices.  The third snapshot illustrates  this situation: there is the same number of type-1  and type-2  vertices. 
\\
(IV) A much slower regime sets in once a FM domain percolates in the $\vec u_\parallel$ direction. This regime is characterised by the emergence of very stable 
FM stripes ($t \gtrsim 10^3$ MCs).  However, the sample is still very far from equilibrium as it needs to grow order in the $\vec u_\perp$ direction as 
well in order to fully equilibrate. The percolation of a domain in this latter direction is achieved by a extremely slow mechanism that we describe below. 
\\
(V) The system equilibrates at much longer times ($t_{eq} \gtrsim 10^{11}$ MCs for this system size).

\begin{figure}[h]
\vspace{0.4cm}
\centering
\includegraphics[scale=.25]{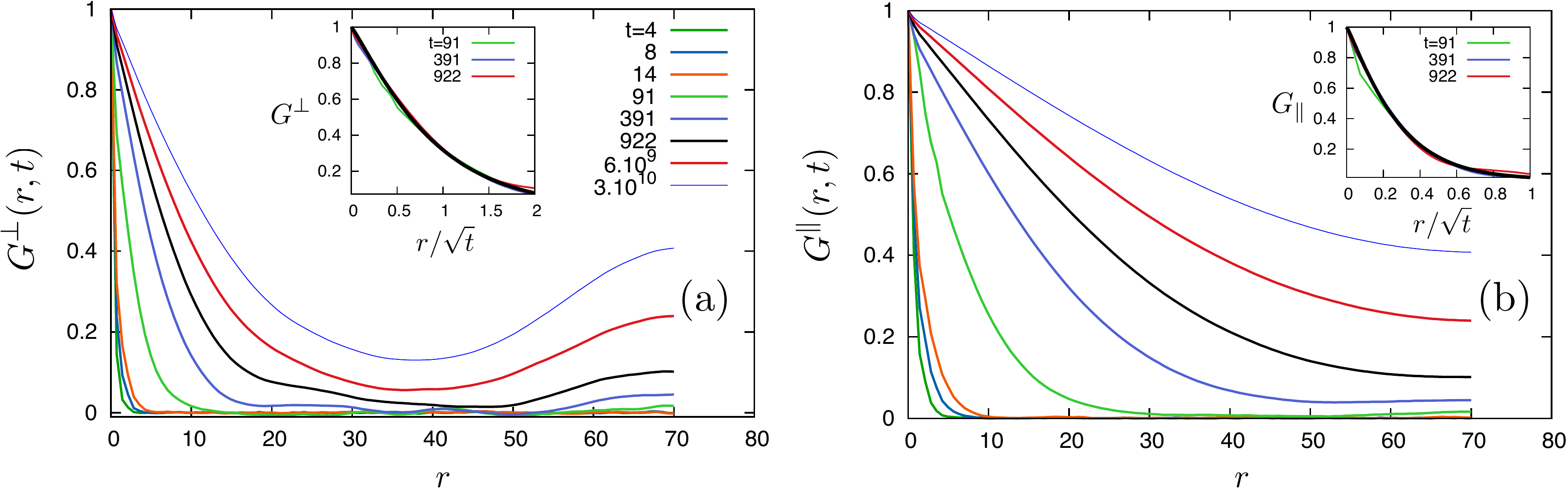} 
\caption{(Colour online.) Quench into the FM phase. 
Space-time correlation function along the orthogonal (a) and longitudinal (b) directions for $L=100$, $a=5$, $b=1$, $d=e^2=10^{-10}$ and different times $t$ 
given in the key. The data have been averaged over $300$ runs. The insets show $G^{\perp,\parallel}$ as a function of the rescaled variable $r/t^{1/2}$ for 
different times in the coarsening regime III. The master curves in the insets are confronted to  $F^{\perp,\parallel}(x)=\exp[-(x/v_{\perp,\parallel})^{w_{\perp,\parallel}})]$ 
where $x=r/\sqrt t$. The scaling functions $F^{\perp,\parallel}$ 
shown in thick black lines were obtained with $v_{\perp}=0.89$, $w_{\perp}=1.15$ (a) and $v_{\parallel}=0.28$, $w_{\parallel}=1.15$ (b).
}
\label{fig:Crt_FM}
\end{figure}

The behaviour of the space-time correlation functions confirm this growth. As shown in Fig.~\ref{fig:Crt_FM} the correlations along the direction parallel to the 
stripes (b) grow faster than in the orthogonal direction (a). The function $G^{\parallel}$ decreases monotonically and does not vanish at any distance for times larger than 
$\approx 10^3$ MCs (regime IV  for which FM stripes percolate). Instead, the correlations along $\vec {u}_\perp$ decrease rapidly at small distances and show a minimum at $L/(2\sqrt{2})$, then $G^{\perp}$ 
increases. This is due to the tendency of the system to develop a modulated structure in time. Because of the periodic boundary conditions a stripe is constrained to wind around the lattice resulting in a 
modulated configuration (as illustrated by  the right-most snapshot in Fig.~\ref{fig:Evolution}). 
The growth is highly anisotropic since $a>b$. For $b>a$ correlations along ${u}_{\perp}$ develop faster than along $\vec {u}_{\parallel}$, forming 
stripes perpendicular to the ones shown in Fig.~\ref{fig:Evolution}. The relevant parameter characterising the anisotropy of the ordering process is the ratio 
$a/b$. As shown in the insets of Fig.~\ref{fig:Crt_FM},  in the regime where anisotropic domains grow (regime III with times $t < 10^3$ MCs)  the correlation function along both directions depends 
on space and time through the ratio $r/t^{1/2}$:
\begin{equation}
G^{\perp,\parallel}(r,t)\simeq F^{\perp,\parallel}\left( \frac{r}{t^{1/2}}\right) \ ,
\end{equation}
which confirms the growing length $L_{\perp,\parallel}(t)\sim t^{1/2}$ found for $d=e$~\cite{Levis2012} with a less refined analysis.
The master curves in the insets of Fig.~\ref{fig:Crt_FM} are confronted to  $F^{\perp,\parallel}(x)=\exp[-(x/v_{\perp,\parallel})^{w_{\perp,\parallel}}]$ where $x=r/\sqrt t$. 
The best fits shown were obtained with $v_{\perp}=0.89$, $w_{\perp}=1.15$ (a) and $v_{\parallel}=0.28$, $w_{\parallel}=1.15$ (b). Notably, the 
stretching exponents $w$ are the same (within numerical accuracy) but the  scales $v_{\perp,\parallel}$ are different.

\begin{figure}[t]
\centering
\includegraphics[scale=1.1]{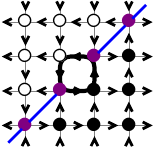}\hspace{0.5cm} 
\includegraphics[scale=1.1]{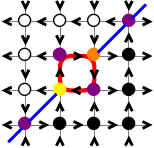} \hspace{0.5cm}
\includegraphics[scale=.12]{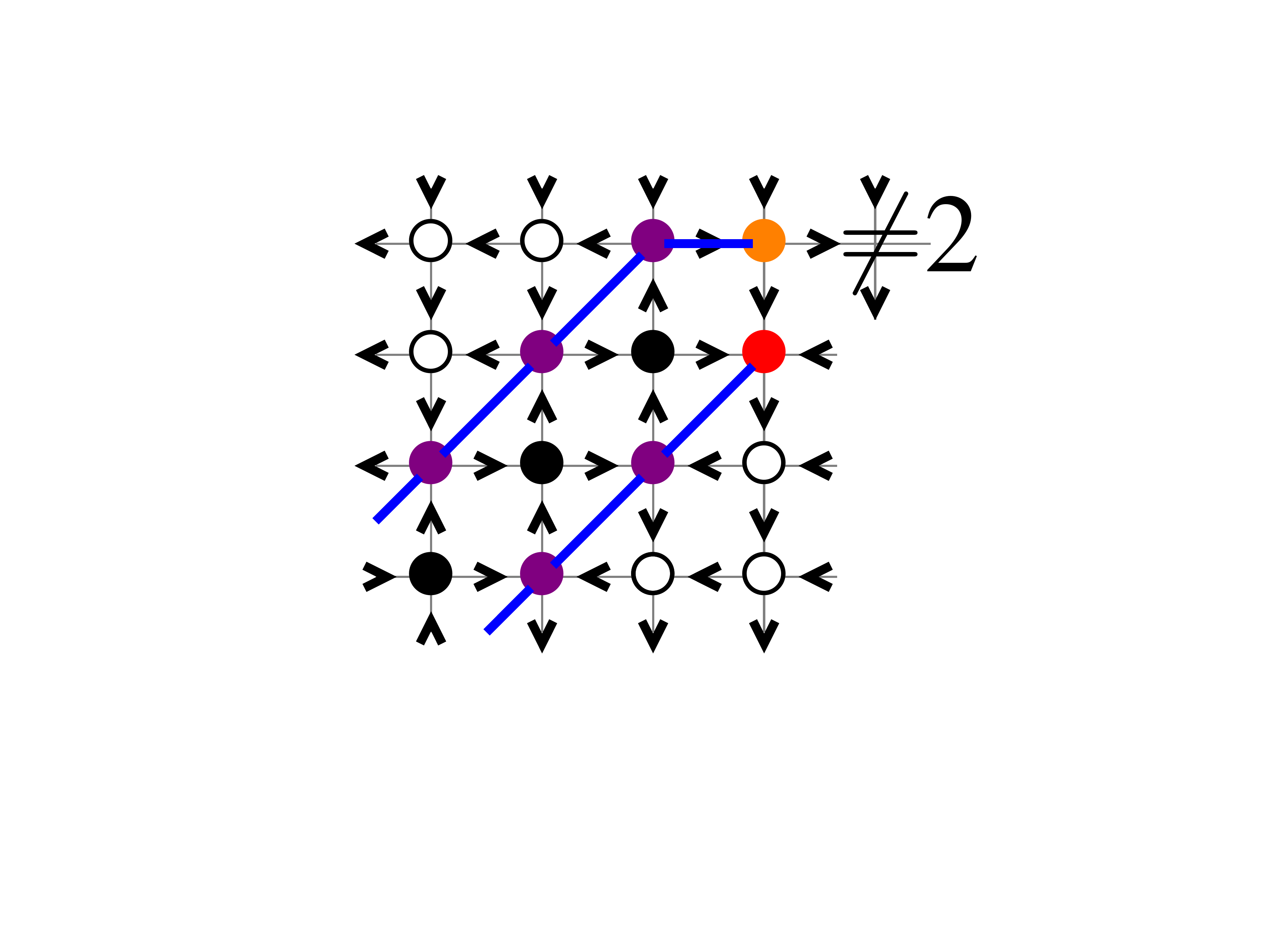} 
\caption{(Colour online.) Interfaces between FM domains. Local
  spins on the bonds and vertices are shown following the Colour rule established in Fig~\ref{vertex_configurations}. Left panel: diagonal
  wall (blue solid line) separating two regions with opposite magnetisation both in the $x$- and $y$-directions.  
  Central panel: a `loop' fluctuation (red solid line) on the plaquette highlighted in the left panel. Right
  panel: a $b$ corner vertex cannot be a neighbour of an $a$-vertex, illustrating the necessity of an outgoing string at the corner of 
  an ordered domain.}
\label{MecanismosDyn}
\end{figure}

\begin{figure}[t]
\centering
\includegraphics[scale=1.1]{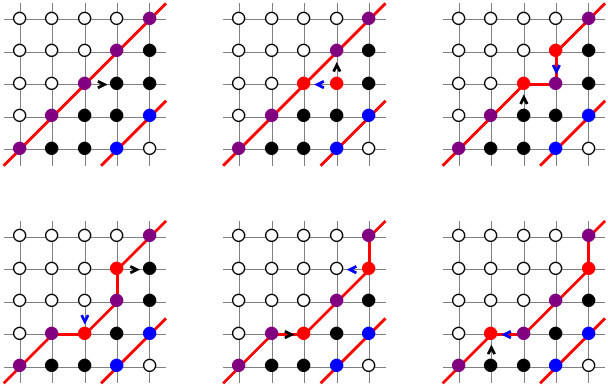} 
\caption{(Colour online.) Schematic representation of FM stripe
  motion. Vertices on each site are specified. Note that the interface is made of 
  antiferromagnetic vertices. Diagonal (red) lines
  delimit domains of opposite magnetisation. Black arrows indicate the
  spins that flip to get the new configuration (represented in blue after the flip).}
\label{MovimientoBandas}
\end{figure}

The dynamical processes leading the dynamics during the different dynamical regimes can be understood from the analysis of the snapshots:

(i) As illustrated in Fig.~\ref{MecanismosDyn}, the topology of the model - i.e. the fact that a vertex of type-$1$ cannot have a neighbour of type-$2$ - leads to 
straight domain walls along the $\vec u_{\parallel}$ direction made by $c$-vertices. Instead, the system would develop interfaces between FM states 
along the  $\vec u_{\perp}$ when quenched into the $b$-favoured FM phase. Plaquettes of ice-rule vertices, as shown in the central panel in 
Fig.~\ref{MecanismosDyn}, can 
frequently appear along the domain walls. The latter are obtained from the elementary excitations of the system. These `loop' fluctuations are obtained by 
sequentially flipping the spins around a plaquette, an operation which preserves the ice-rule (see the first and second panel in Fig.~\ref{MecanismosDyn}). 

(ii)  Domains of the same type are connected by quasi-one-dimensional
paths made of $b$- and $c$-vertices (loop fluctuation can eventually be attached
to them as well) running through a region with the opposite order. 
The interplay between the tendency to order  and the local constraint gives rise to these structures. 
They are similar to the ones found in the kinetically constrained spiral
model~\cite{Corberi2009}. 
 In order to further increase the density of $a$-vertices
and develop the FM order, the domain
walls and strings of $b$- and $c$-vertices have to be eliminated.  
The latter disappear first via the following mechanism.
Curved domains must have `corners' made of $b$ vertices or defects, but 
$b$ vertices on the corners cannot be surrounded by more
than two type 1 or 2 vertices (only defects can, 
giving rise to the before mentioned quasi one-dimensional structures, as illustrated in the third panel
in Fig.~\ref{MecanismosDyn}).  The string then progressively disappears
eaten by the attached domains that grow from the corner or,
alternatively, it is first cut by the creation of two defects and the
two strands subsequently shrink. 
Once the string has been eliminated one is left with two defects sitting on the
walls of the now detached domains, that move along the interface and
eventually annihilate with their anti-partner. 

(iii) Once the domains have grown enough to percolate in the $\vec u_\parallel$ direction the mechanism sketched in 
Fig.~\ref{MovimientoBandas} takes over. 
The formation of stable parallel stripes freezes the dynamics. The only way to evolve from a configuration with parallel FM stripes is to create a 
pair of defects by a single spin-flip.  After the creation of a pair of defects on the domain wall, the diffusion of the latter along the 
interface shrink one among the two oppositely ordered stripes. This is illustrated by the sequence of steps in 
Fig.~\ref{MovimientoBandas}. The number of steps scales as the length of the 
stripe, hence the time associated to this process diverges with the size of the system.

\subsection{Isotropic $c$-AF domain growth}
\label{subsec:coarsening-AF}

We now turn to quenches into the AF phase. To start with we present results for parameters that favour isotropy, that is to say
$a=b$, and defect weights such that $e>d$. This is the case realised in as-grown ASI samples~\cite{Morgan2011,Morgan2013}. 
Next we show some snapshots for non-equal FM weights, $a\neq b$, 
though still in the AF phase to demonstrate that anisotropic AF growth is also possible.

\subsubsection{$c$-AF sixteen-vertex model with $d=e^2$}

\begin{figure}[h]
\centering
 \includegraphics[scale=0.38]{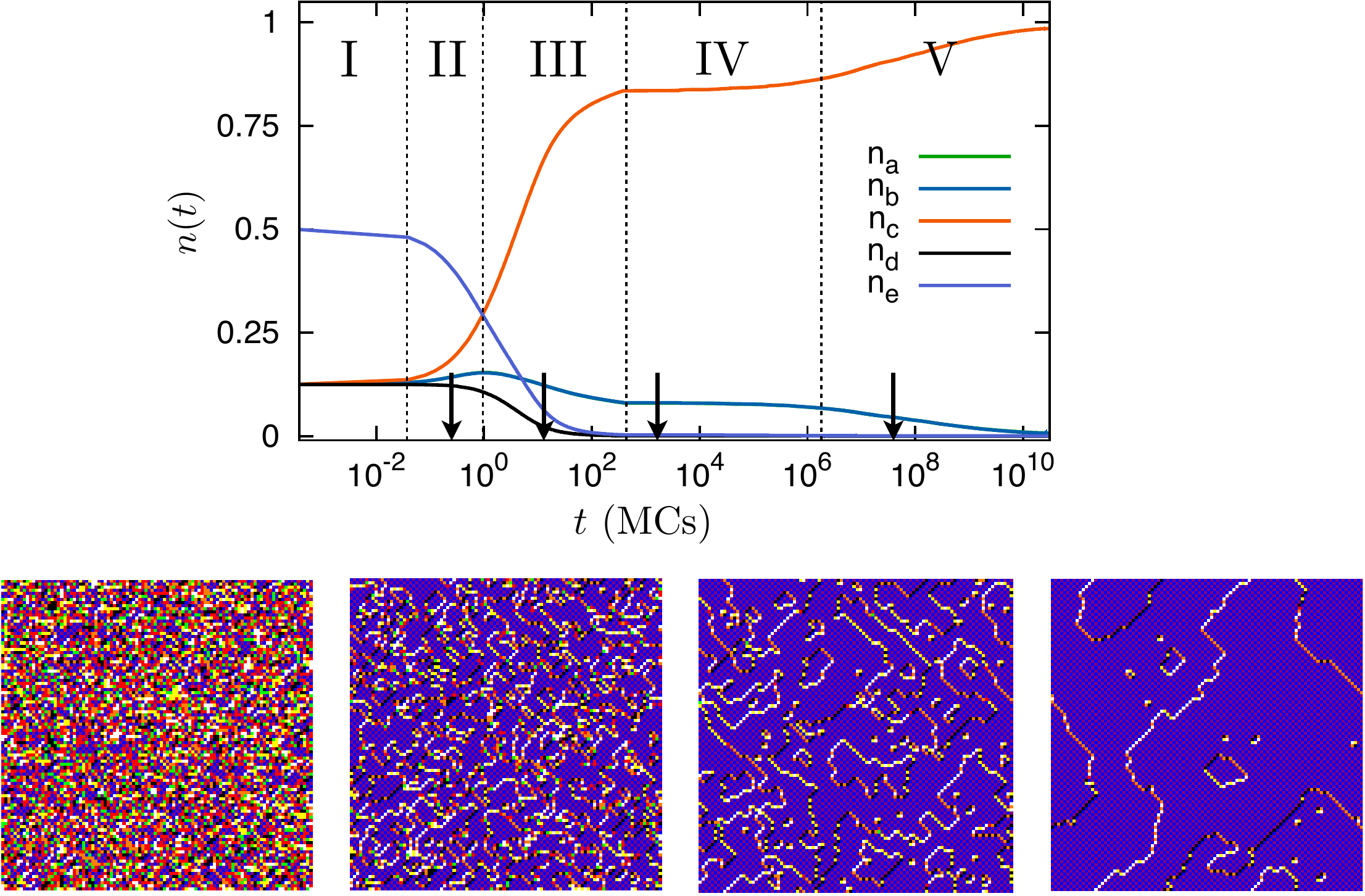}
\caption{(Colour online.) $c$-AF ordering. Time evolution of the density of vertices in a system with 
$L=100$ after a quench to $a=0.1$, $b=0.1$, $d=e^2=10^{-10}$ averaged over 500 runs. Snapshots taken at the instants indicated by arrows in the main 
panel are shown (using the colour code defined in Fig.~\ref{vertex_configurations}).}
\label{fig:EvolutionAF}
\end{figure}

In Fig.~\ref{fig:EvolutionAF} we show the evolution of the vertex populations, $n_{\kappa}$, following a quench into the $c$-AF phase 
by setting $a=b=10^{-1}$ and 
$d=e^2=10^{-10}$.  This data are illustrated by configurations taken at different instants indicated by arrows in the main panel. This choice of 
parameters favours $c$-vertices and is invariant under $\pi/2$-rotations. Therefore, the evolution proceeds by growing isotopic domains of opposite 
staggered magnetisation denoted by $m_{-}^{x,y}=\pm1$ (the staggered orientation of the spins lying on horizontal and vertical edges respectively, see 
Fig.~\ref{fig:CorrLattice}). 

\begin{figure}[h]
\vspace{0.4cm}
\centering
 \includegraphics[scale=0.25]{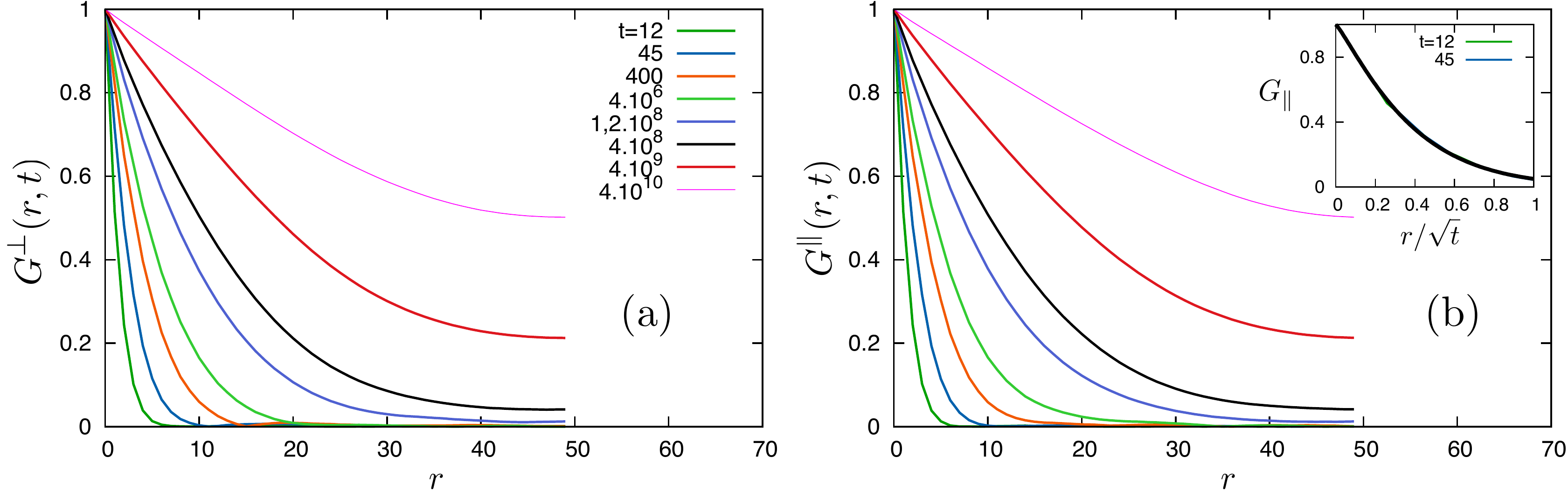}
\caption{(Colour online.) Quench into the $c$-AF phase.
Space-time correlations after a quench from a random initial condition into  $a=0.1=b=0.1$, $d=e^2=10^{-10}$  and $L=50$ averaged over 500 runs. (a) 
Correlations along the $\perp$ direction as a function of the distance $r$ between sites for different times (shown in the key). (b) Correlations along the 
$\parallel$ direction as a function of  $r$ for the same times as in (a).  The inset shows the scaling of $G^{\parallel}$ 
as a function of $r/\sqrt{t}$ confronted to $g(x)=\exp[-(x/v)^{w})]$ with $x=r/\sqrt t$ (shown in thick black lines). The best fit shown 
in the inset was obtained with  $v=0.39$ and  $w=1.17$, that is very close to the value found for the stretching exponent in the FM phase. }
\label{fig:CrtAF}
\end{figure}

With a similar analysis to the one used for the FM quench, different dynamical regimes can be identified:
\\
(I) A short transient where all the densities remain roughly constant. 
\\
(II) An intermediate regime ($t \lesssim 10$ MCs) with a rapid annihilation of defects into ice-rule vertices. 
\\
(III) A third regime during which $n_a=n_b$ decreases in order to grow  isotropic $c$-ordered domains. The identity of $a$ and $b$ vertex 
weights implies that $n_a$ and $n_b$ are equal. This is explicitly shown by the data plotted in 
Fig.~\ref{fig:CrtAF}. The space-time self correlation functions along the $\parallel$ and $\perp$ directions are almost identical 
and the associated growing lengths are, within numerical accuracy, $t^{1/2}$. The scaling of the space-time correlation
function shown in the inset to panel (b) confirms this claim. Moreover, the scaling function is again a stretched exponential with an
stretching exponent $w$ that is very close to the one found in the FM quenches. The scale $v$ depends on the working parameters.
For these set of parameters and system size, 
regime III is relatively short, it lasts until $t\simeq 5 \ 10^2$ MCs.
\\
(IV) Although we have not performed a full statistical analysis of this fact, it seems to us that the system freezes when an ordered structure 
winds around the finite size sample for periodic boundary conditions (or goes from one border to another for open boundary conditions).
This regime is the one in which the plateau in the density of defects, discussed in the previous Section, emerges. In the plot in which we show the 
densities of each kind of vertices separately one sees that all densities are constant in this regime. The process whereby the system leaves this regime will be 
discussed below. The entrance into this regime will take longer times for larger sizes as already discussed in Sec.~\ref{sec:Metastability}. 
\\
(V) At longer times the system finally reaches equilibrium. 

A better understanding of the processes involved in the ordering dynamics 
is reached from the analysis of 
the snapshots. 

\begin{figure}[h]
\vspace{0.2cm}
\centering
\includegraphics[scale=.3]{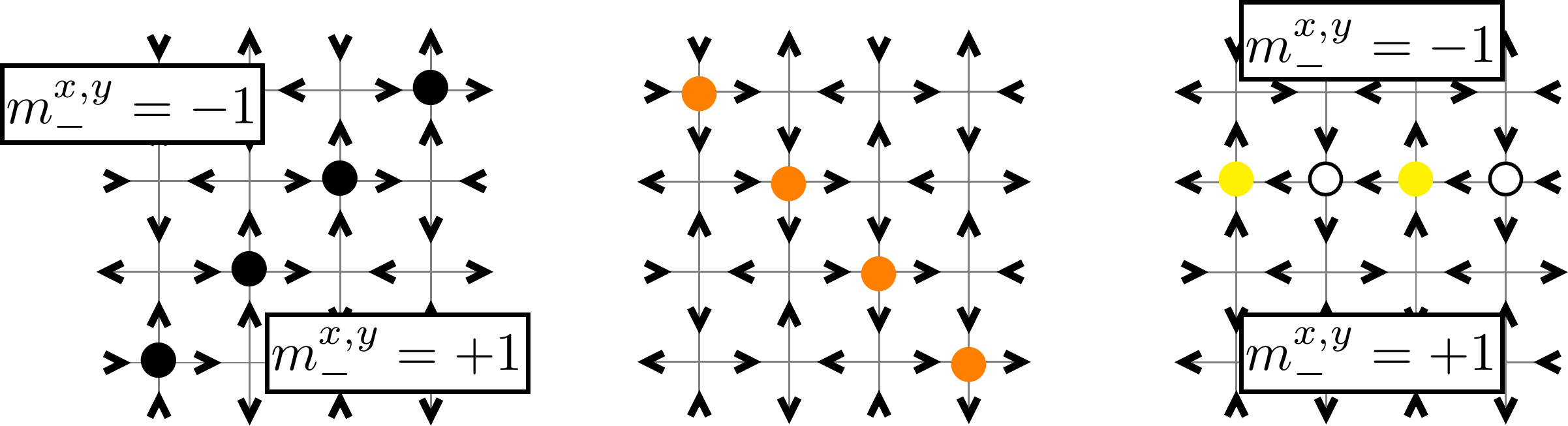} 
\caption{(Colour online.) Domain walls between $c$-AF domains of opposite staggered order. The configuration of arrows is shown. We use the colour rule defined in Fig.~\ref{vertex_configurations}. 
Left panel: diagonal walls in the $\parallel$ direction are made of $a$-vertices. Central panel: diagonal walls in the $\perp$ direction are made of $b$-vertices. Right panel: horizontal 
(and vertical, by symmetry) walls are made of an alternating chain of $a$- and $b$-vertices. }
\label{fig:DomainAF}
\end{figure}

(i) Domain walls are made of
$a$- and $b$-vertices. Contrarily to the FM case, domains of any shape can be constructed without defects. As shown in
the left and central panels in Fig.~\ref{fig:DomainAF}, horizontal and vertical walls are made by alternating $a$- and $b$-vertices. Diagonal 
walls are exclusively made by $a$- or $b$-vertices depending on their orientation.
Therefore, domain walls without defects (energetically favoured) form loops of spins pointing along the same direction. 

Dynamic domain walls in, say, Ising magnetic models with non-conserved order parameter (NCOP) 
dynamics are curved, and at finite times they display a variety 
of shapes, are relatively smooth at short length-scales scales and have fractal properties at long length-scales~\cite{Arenzon2007,Sicilia2007}. Instead, in the sixteen-vertex model
with $d=e^2\ll 1$ $c$-AF domains have tendency to form straight walls made by FM 
vertices as shown in the right most snapshot in Fig.~\ref{fig:EvolutionAF}.  A statistical and geometric analysis of 
the morphology of domains and interfaces and their dependence on the parameters of the model 
remains an interesting project, especially if one wishes to confront the predictions of this model 
to images of artificial spin ice samples.

(ii) In the Solid-on-Solid (SOS) representation of the six-vertex model~\cite{Beijeren1977}, each domain can be interpreted  as a contour line 
delimiting regions with different height. The ordering then proceeds by growing or shrinking regions of constant height. Note that the Kosterlitz-Thouless phase 
transition of the F-model ($a=b$) can be mapped onto the roughening transition~\cite{Beijeren1977}.  The ordering dynamics in this phase thus 
correspond to flatten the initial rough surface by growing regions made by $c$-vertices. 

\begin{figure}[h]
\vspace{0.4cm}
\centering
\includegraphics[scale=.3]{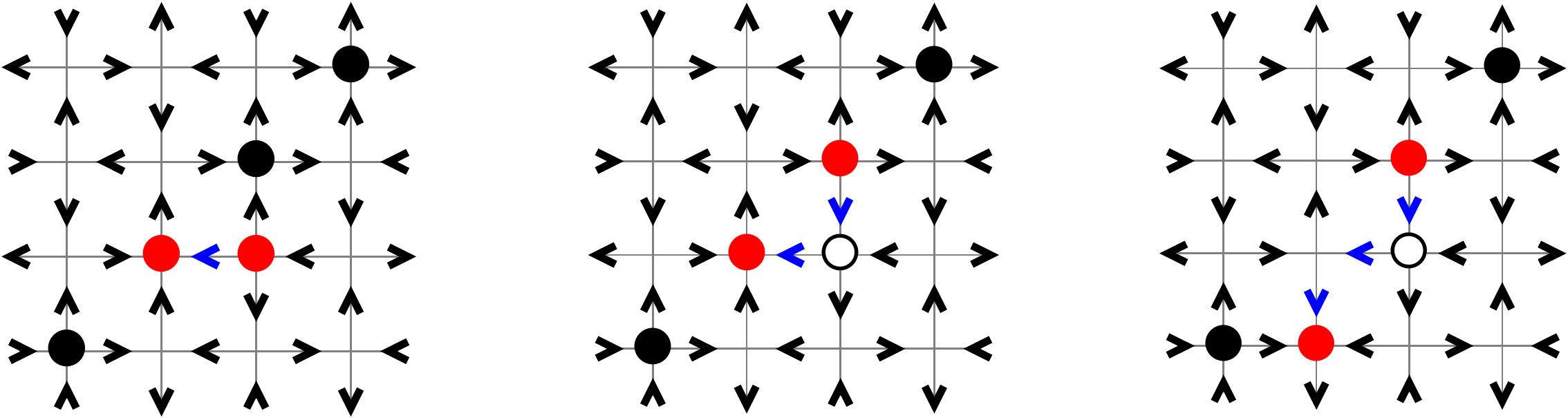} 
\caption{(Colour online.) Schematic representation of the annihilation of $c$-AF domains. Vertices on the walls are represented by the colour rule defined in Fig.~\ref{vertex_configurations}.  
Blue arrows indicate the
  spins that have been flipped to get the new configuration. As defects diffuse along the domain wall the ordered region on the right-bottom side of the figure shrinks. }
\label{fig:MecaAF}
\end{figure}

(iii) Once isotropic domains are created, one has to eliminate small domains in order to further increase the density of $c$-vertices {of the 
losing kind}
and develop the {conquering} $c$-AF order. Figure~\ref{fig:MecaAF} illustrates the mechanism taking place. After the creation of a pair of 
defects in a typical time 
$\sim 1/e^2$, their motion along the wall shrinks the domain.  This is done without any energy cost and the sequence of steps needed to make 
a domain  disappear should scale with its size.  The same kind of mechanism takes place for horizontal domain walls.

\subsubsection{Artificial Spin Ice}

As shown in~\cite{Levis2013}, one can choose the vertex weights $c>a=b>e>d$ in order to model as-grown ASI samples~\cite{Morgan2011,Morgan2013}.   
In these experiments, ferromagnetic islands are gradually grown by deposition. During the early stages of the preparation the system feels thermal 
fluctuations and thus tries to accommodate to its $c$-AF ground state. Once the islands reach a critical size the system blocks into a frozen configuration. The 
question as to whether such frozen configurations sample the equilibrium distribution has been recently addressed 
experimentally~\cite{Morgan2011, Morgan2013} and theoretically~\cite{Levis2013}.
Using this latter choice of parameters inspired by ASI samples:  $a=\exp({-\beta\epsilon_a})$, $b=\exp({-\beta\epsilon_b})$, 
 $c=\exp({-\beta\epsilon_c})$, $d=\exp({-\beta\epsilon_d})$ and 
$e=\exp(-\beta\epsilon_e)$, with $\epsilon_a=\epsilon_b=2/l$, $\epsilon_c=2(1-2\sqrt{2})/l$, $\epsilon_d=2(2\sqrt2 +1)/l$ and 
$\epsilon_e=0$ ($l$ being the lattice constant); 
 the AF domain walls become smoother than the ones obtained for $d=e^2=10^{-10}$. It is 
due to the presence of more defects lying on the interfaces (see Fig. 6 in~\cite{Levis2013}). 
This kind of domain wall pattern has also been observed in as-grown artificial spin-ice samples~\cite{Morgan2011}. {This is why we 
believe that some of these samples, the ones that are close to the phase transition, are out of equilibrium 
although the density of defects is notably well described by the model in equilibrium~\cite{Foini2012}}
and in simulation studies of a spin-ice model system~\cite{Budrikis2012}.

\begin{figure}[h]
\vspace{0.4cm}
\centering
\includegraphics[scale=.4]{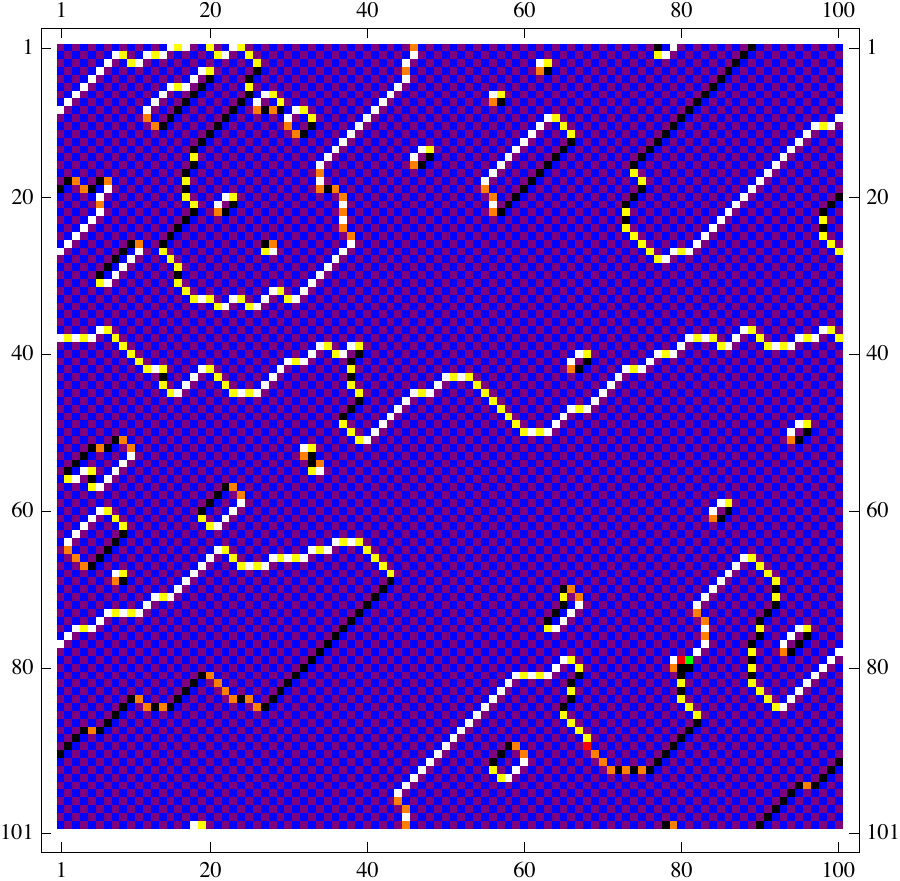} \hspace{1cm}
\includegraphics[scale=.4]{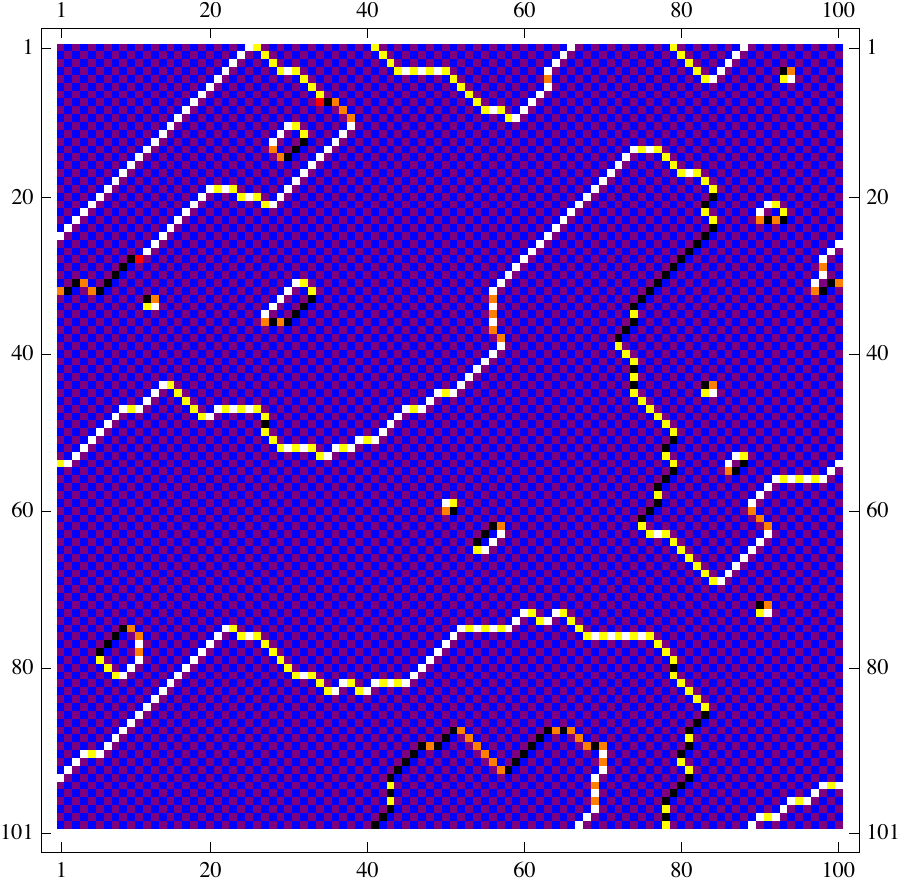}
\caption{(Colour online.) Typical configurations of  an $L=100$ lattice with $a=0.1$, $b=0.01$, $d=e^2=10^{-10}$ (left) and $a=0.1$, 
$b=0.001$, $d=e^2=10^{-10}$ (right). The two snapshots have been taken at different times during the ordering regime IV in the $c$-AF phase. } 
\label{fig:AFsnapAnis}
\end{figure}

Therefore, the shape of the interfaces separating opposite AF regions is extremely sensitive to the choice of parameters. Setting $a$ slightly above (below) 
$b$ would produce a preference to grow domains in the $\vec u_{\parallel}$ ($\vec u_\perp$) direction. This is illustrated by the two 
snapshots shown in Fig.~\ref{fig:AFsnapAnis}.
As shown in~\cite{Foini2012}, the equilibrium fluctuations of 
the model reflect this feature as well.  
The evolution of the system after a quench into the $c$-AF phase would eventually get frozen into an extremely slow relaxation regime, due to the presence of 
percolating domains in the $\vec u_{\parallel}$ ($\vec u_\perp$) direction. The anisotropic dynamical scaling is difficult to study with our numerical methods in this 
case, and it is not clear whether the same scaling ($L_{\parallel}\sim t^{1/2}$) holds along both directions. Experimentally, the anisotropy ratio $a/b$ can be 
tuned by placing islands of different length along the horizontal and vertical edges. This work should incite experimentalists to study this rich far from 
equilibrium behaviour in ASI.    

\section{Conclusion}
\label{sec:conclusions}

In this paper we presented a thorough study of the relaxational stochastic dynamics of the sixteen vertex model with low defect weights
after quenches from infinite temperature to the three equilibrium phases: disordered, ferromagnetic and antiferromagnetic ones.

In~\cite{Levis2012,Foini2012,Levis2013} we proposed that this celebrated model captures
many important aspects of two dimensional artificial spin ice systems. The interest in these materials lies on the fact that they 
are a possible new technology for storage data devices. Our approach to these systems is, clearly, theoretical as their fabrication and preparation 
pose a number of rather fundamental questions that we can try to give an answer to by better understanding the behaviour of the model.

In the present publication we showed that, for sufficiently low defect weight and finite though large system size, the single spin-flip stochastic dynamics 
of the model that mimics thermal agitation in artificial spin ice,  after quenches from infinite temperature, present several, very distinct, collective 
dynamic regimes. In short, these regimes are the rapid annihilation of defects at very short time scales, coarsening of nearly critical or
phase ordering kind in the paramagnetic or ordered phases, respectively, metastability over a very long period of time, and 
the final approach to equilibrium. The coarsening process in the paramagnetic phase is controlled by the proximity of the critical 
phase in the absence of defects, the so-called spin-liquid phase of the six vertex model. The phase ordering kinetics in the 
ferromagnetic phases is characterised by anisotropic growth of the ferromagnetic equilibrium states related by spin-reversal. The 
ordering process in the anti-ferromagnetic phase is isotropic for parameters belonging to the so-called F-model ($a=b$) but it is not 
isotropic for other choices of weights.

In quenches to the FM phase we found that the end of the first coarsening regime is attained when two diagonal stripes of opposite 
ferromagnetic order percolate across the sample. Note that these states are equivalent from a thermodynamic point of view
and no interaction in the model's Hamiltonian distinguishes between them.
It is the dynamics that develops this transient, although very long-lived, anisotropy. The crossover from coarsening to metastability clearly depends on the 
size of the sample and the growth law $t^{1/2}$ suggests a crossover-time $t_{co} \simeq L^2$ when an ordered band goes from one system 
border to another. 
The fact that the dynamics get blocked with {\it two} stripes is based 
on the visual analysis of the snapshots and the fact that the perpendicular space-time correlation reaches a minimum at a distance of the order 
of $L/(2\sqrt{2})$. A more refined analysis along the lines in~\cite{Krapivsky,Blanchard}  is necessary to 
clarify the role played by the initial condition in the number of stripes in the blocked state. As the time needed to let the stripes percolate
across the sample diverges with the system size, it is clear that metastability and the further approach to equilibrium (regimes IV and V) 
are pushed to infinity in the thermodynamic limit.

In quenches to the $c$-AF phase the entrance in the metastable regime is also achieved when the two domains with the opposite staggered 
magnetisation percolate
and, therefore, the last two regimes are moved to divergent time-scales in the thermodynamic limit in this phase as well. AF ordering 
is characterised by a $t^{1/2}$ growing length.

In quenches to the PM phase, the proximity to the critical spin-liquid of the six vertex model leads to a  relaxation with 
a very long relaxation time that diverges in the limit of zero defect weight. Equilibrium patches are hard to visualise in a 
model with many local variables as the single vertices we have in this model but, most probably, the cross-over to 
metastability is also controlled  by percolation of one (or more) of these regions in this phase. A critical growing 
length that should saturate to a finite value should exist in this regime as well.

We found stretch exponential scaling functions for the space-time correlation functions in the two anisotropic directions during FM coarsening and  
in the isotropic AF coarsening. The fits are consistent with the same (within numerical accuracy) stretching exponent $w$ but different 
scale $v$, i.e. for a scaling function $F(x) \simeq \exp[-(x/v)^w]$.

Long-lived metastable states exist in the three phases after a cross-over time that should diverge with the system size. Still, 
as artificial spin-ice samples are of finite size, we studied these states in detail here. 
The configurations found in these time scales, that we called regime IV  in the Sections concerning the ordered phases,
are characterised by densities of each kind of vertex that are 
still far from their equilibrium values for the parameters that we investigated in this paper (note, however, that agreement between 
dynamic and static densities of vertices can be achieved for other parameters, as some of the ones used in~\cite{Foini2012}
with the aim of confronting to experimental data). In particular, we found here a finite number of 
defects in the dynamic configurations. Our finite size scaling analysis suggests that 
their density vanish in the infinite size limit though it is hard to reach a definite conclusion in this respect. 

Regime IV is succeeded by a totally different epoch during 
which complete ordering is eventually reached. Several mechanisms for the motion of domain walls 
in the FM and AF phases that constitute the relevant relaxation processes in this time regime
have been identified and discussed in the text.

Let us conclude by confronting our approach to other dynamic studies of the same 
physical problem, reported in the literature in the last two or three years.
Metastability in the defect density of frustrated magnets was discussed in~\cite{Castelnovo2010}
by using a reaction-diffusion model in which the defects are represented by charges and interactions of Coulomb type 
are considered. In this modelling the `background configuration'  is not taken into account. In our model, instead, 
the structure in which the defects are placed and displace is very important and determines the crossover time 
to metastability, the frozen nature of the system, the escape time from this frozen state and the subsequent final 
relaxation to equilibrium.
 In~\cite{Budrikis2012}, a mean-field model of spin-ice, with only short range interactions, is proposed in order to study the domain wall dynamics of ASI in the presence of disorder and magnetic fields,
 two `external' perturbations that we have not taken into account in our work. 
A very different approach has been taken in~\cite{Wysin2013} to address spin-ice dynamics in $2D$. In this work the authors introduce a model of continuous magnetic moments with dipolar interactions and 
study its stochastic dynamics. The strong anisotropy imposed along the edges of the underlying square lattice makes the connection with ASI samples possible. 
From a more general perspective on $2D$ frustrated magnetism, numerical studies of the constrained dynamics of a simple classical lattice model have shown the existence of a dynamical arrest analogous to 
the one discussed in the text~\cite{Cepas2012}. In this work the defects breaking a local constraint (analogous to the ice-rules) are strictly forbidden contrarily to 
what we do in our paper. Our work presents a new approach that allows to deal with  
thermal fluctuations breaking a hard constraint, and its consequences in the out-of-equilibrium relaxation of geometrically frustrated magnets.

\bibliographystyle{phjcp}
\bibliography{16V_Dynamics-short}

\begin{thebibliography}{10}

\bibitem{Balents2010}
{\sc L.~Balents},
\newblock {\em Nature} {\bf 464}, 199 (2010).

\bibitem{Wang2006}
{\sc R.~F. Wang}, {\sc C.~Nisoli}, {\sc R.~S. Freitas}, {\sc J.~Li}, {\sc
  W.~McConville}, {\sc B.~J. Cooley}, {\sc M.~S. Lund}, {\sc N.~Samarth}, {\sc
  C.~Leighton}, {\sc V.~H. Crespi}, and {\sc P.~Schiffer},
\newblock {\em Nature} {\bf 439}, 303 (2006).

\bibitem{Nisoli2007}
{\sc C.~Nisoli}, {\sc R.~Wang}, {\sc J.~Li}, {\sc W.~McConville}, {\sc
  P.~Lammert}, {\sc P.~Schiffer}, and {\sc V.~Crespi},
\newblock {\em Phys. Rev. Lett.} {\bf 98}, 217203 (2007).

\bibitem{Morgan2011}
{\sc J.~P. Morgan}, {\sc A.~Stein}, {\sc S.~Langridge}, and {\sc C.~H.
  Marrows},
\newblock {\em Nature Phys.} {\bf 7}, 75 (2010).

\bibitem{Cowburn-Welland}
{\sc R.~P. Cowburn} and {\sc M.~E. Welland},
\newblock {\em Science} {\bf 287}, 1466 (2000).

\bibitem{Bader}
{\sc S.~D. Bader},
\newblock {\em Rev. Mod. Phys.} {\bf 78}, 1 (2006).

\bibitem{Harris1997}
{\sc M.~J. Harris}, {\sc S.~T. Bramwell}, {\sc D.~F. Mcmorrow}, {\sc
  T.~Zeiske}, and {\sc K.~W. Godfrey},
\newblock {\em Phys. Rev. Lett.} {\bf 79}, 2554 (1997).

\bibitem{DiepBookCH7}
{\sc S.~T. Bramwell}, {\sc M.~J.~P. Gingras}, and {\sc P.~C.~W. Holdsworth},
\newblock in {\em Frustrated spin systems}, edited by {\sc H.~T. Diep}, World
  Scientific, 2004.

\bibitem{Moller2006}
{\sc G.~M\"oller} and {\sc R.~Moessner},
\newblock {\em Phys. Rev. Lett.} {\bf 96}, 237202 (2006).

\bibitem{Wysin2013}
{\sc G.~M. Wysin}, {\sc W.~A. Moura-Melo}, {\sc L.~A.~S. M\'ol}, and {\sc A.~R.
  Pereira},
\newblock {\em arXiv:1208.6557}  (2012).

\bibitem{Levis2012}
{\sc D.~Levis} and {\sc L.~F. Cugliandolo},
\newblock {\em EPL} {\bf 97}, 30002 (2012).

\bibitem{Foini2012}
{\sc L.~Foini}, {\sc D.~Levis}, {\sc M.~Tarzia}, and {\sc L.~F. Cugliandolo},
\newblock {\em J. Stat. Mech} , P02026 (2013).

\bibitem{BaxterBook}
{\sc R.~J. Baxter},
\newblock {\em {Exactly solved models in statistical mechanics}},
\newblock Dover, 1982.

\bibitem{LiebWuBook}
{\sc E.~H. Lieb} and {\sc F.~Y. Wu},
\newblock in {\em Phase transitions and critical phenomena Vol. 1}, edited by
  {\sc C.~Domb} and {\sc J.~L. Lebowitz}, chapter~8, Academic Press, 1972.

\bibitem{Baxter1971}
{\sc R.~J. Baxter},
\newblock {\em Phys. Rev. Lett.} {\bf 26}, 832 (1971).

\bibitem{Budrikis2012}
{\sc Z.~Budrikis}, {\sc K.~L. Livesey}, {\sc J.~P. Morgan}, {\sc J.~Akerman},
  {\sc A.~Stein}, {\sc S.~Langridge}, {\sc C.~H. Marrows}, and {\sc R.~L.
  Stamps},
\newblock {\em New J. Phys.} {\bf 14}, 035014 (2012).

\bibitem{Bernal1933}
{\sc J.~D. Bernal} and {\sc R.~H. Fowler},
\newblock {\em J. Chem. Phys.} {\bf 1}, 515 (1933).

\bibitem{Pauling1935}
{\sc L.~Pauling},
\newblock {\em J. of Chem. Phys.} {\bf 57}, 2680 (1935).

\bibitem{Levis2013}
{\sc D.~Levis}, {\sc L.~F. Cugliandolo}, {\sc L.~Foini}, and {\sc M.~Tarzia},
\newblock {\em arXiv:1302.3725}  (2013).

\bibitem{Ryzhkin2005}
{\sc I.~A. Ryzhkin},
\newblock {\em J. Exp. Theor. Phys.} {\bf 101}, 481 (2005).

\bibitem{Castelnovo2008}
{\sc C.~Castelnovo}, {\sc R.~Moessner}, and {\sc S.~Sondhi},
\newblock {\em Nature} {\bf 451}, 42 (2008).

\bibitem{BKL}
{\sc A.~B. Bortz}, {\sc M.~H. Kalos}, and {\sc J.~L. Lebowitz},
\newblock {\em J. Comp. Phys.} {\bf 17}, 10 (1975).

\bibitem{Chakraborty2002}
{\sc B.~Chakraborty}, {\sc D.~Das}, and {\sc J.~Kondev},
\newblock {\em Eur. Phys. J. E} {\bf 9}, 227 (2002).

\bibitem{Cepas2012}
{\sc O.~Cepas} and {\sc B.~Canals},
\newblock {\em Phys. Rev. B} {\bf 86}, 024434 (2012).

\bibitem{Korepin-ZinnJustin}
{\sc V.~Korepin} and {\sc P.~Zinn-Justin},
\newblock {\em J. Phys. A} {\bf 33}, 7053 (2000).

\bibitem{Zinn-Justin}
{\sc P.~Zinn-Justin},
\newblock {\em Phys. Rev. E} {\bf 62}, 3411 (2000).

\bibitem{Castelnovo2010}
{\sc C.~Castelnovo}, {\sc R.~Moessner}, and {\sc S.~L. Sondhi},
\newblock {\em Phys. Rev. Lett.} {\bf 104}, 107201 (2010).

\bibitem{Morgan2013}
{\sc J.~P. Morgan}, {\sc J.~Akerman}, {\sc A.~Stein}, {\sc C.~Phatak}, {\sc
  R.~M.~L. Evans}, {\sc S.~Langridge}, and {\sc C.~H. Marrows},
\newblock {\em Phys. Rev. B} {\bf 87}, 024405 (2013).

\bibitem{Corberi2009}
{\sc F.~Corberi} and {\sc L.~F. Cugliandolo},
\newblock {\em J. Stat. Mech.} , P09015 (2009).

\bibitem{Arenzon2007}
{\sc J.~J. Arenzon}, {\sc A.~J. Bray}, {\sc L.~F. Cugliandolo}, and {\sc
  A.~Sicilia},
\newblock {\em Phys. Rev. Lett.} {\bf 98}, 145701 (2007).

\bibitem{Sicilia2007}
{\sc A.~Sicilia}, {\sc J.~J. Arenzon}, {\sc A.~J. Bray}, and {\sc L.~F.
  Cugliandolo},
\newblock {\em Phys. Rev. E} {\bf 76}, 061116 (2007).

\bibitem{Beijeren1977}
{\sc H.~van Beijeren},
\newblock {\em Phys. Rev. Lett} {\bf 38}, 993 (1977).

\bibitem{Krapivsky}
{\sc K.~Barros}, {\sc P.~L. Krapivsky}, and {\sc S.~Redner},
\newblock {\em Phys. Rev. E} {\bf 80}, 040101 (2009).

\bibitem{Blanchard}
{\sc T.~Blanchard} and {\sc M.~Picco},
\newblock {\em arXiv:}  (2013).

\end{thebibliography}
\end{document}